\documentclass{ailab}
\usepackage{lmodern}
\usepackage{amsmath,amsfonts,bm}

\def\eqref#1{equation~\ref{#1}}

\def\1{\bm{1}}

\DeclareMathAlphabet{\mathsfit}{\encodingdefault}{\sfdefault}{m}{sl}
\SetMathAlphabet{\mathsfit}{bold}{\encodingdefault}{\sfdefault}{bx}{n}

\usepackage{algorithm}
\usepackage{algpseudocode}
\usepackage{float}
\usepackage{comment}
\usepackage{url}
\usepackage{ragged2e}
\graphicspath{{figures/}}

\title{\centering Self-Supervised Scaling of Terminal Environments for Scientific Domains}

\author[1,2]{Zhongzhi Li\textsuperscript{*}}
\author[1]{Yucheng Shi\textsuperscript{*}\textsuperscript{\(\dagger\)}}
\author[1,3]{Zongxia Li\textsuperscript{*}}
\author[1,4]{Junyao Yang}
\author[1,5]{Ruhan Wang}
\author[6]{Yu Wang}
\author[2]{Jingyuan Huang}
\author[2]{Jichao Yu}
\author[1]{Haitao Mi}
\author[3]{Jing Xie}
\author[3]{Xiaomin Wu}
\author[7]{Ninghao Liu\textsuperscript{\(\ddagger\)}}
\author[1]{Leowei Liang}

\affiliation[1]{Tencent HY LLM Frontier}
\affiliation[2]{University of Georgia}
\affiliation[3]{University of Maryland, College Park}
\affiliation[4]{National University of Singapore}
\affiliation[5]{Indiana University}
\affiliation[6]{University of Illinois at Chicago}
\affiliation[7]{Hong Kong Polytechnic University}

\contribution{\textsuperscript{*}Equal contribution}
\contribution{\textsuperscript{\(\dagger\)}Project lead.}
\contribution{\textsuperscript{\(\ddagger\)}Corresponding author.}

\email{ninghliu@polyu.edu.hk, zl22754@uga.edu, zhongzhili@global.tencent.com}

\newcommand{\opub}{O_{\mathrm{pub}}}
\newcommand{\spub}{S_{\mathrm{pub}}}
\newcommand{\shid}{S_{\mathrm{hid}}}
\newcommand{\judge}{J}
\newcommand{\instr}{I}
\newcommand{\gtheta}{g_{\theta}}
\newcommand{\rpub}{r_{\mathrm{pub}}}
\newcommand{\rhid}{r_{\mathrm{hid}}}

\abstract{%
Terminal agents are increasingly deployed beyond software engineering in science and other specialized domains. Constructing training environments in these domains requires executable reference solutions and domain-specific verifiers that check whether an agent's outputs are correct. Authoring these components independently for every domain requires repeated engineering effort and limits reuse across tasks. We introduce \emph{software-in-the-loop reconstruction}, a self-supervised learning framework that uses existing software workflows as reference solutions and generates reference outputs by executing them. For each workflow, we execute multiple input configurations and split the resulting input--output pairs into public examples and hidden test cases. Given the task instruction and public input--output examples, an agent constructs a candidate program to reproduce the workflow's computational behavior without access to its reference implementation. We evaluate the candidate program on hidden inputs by comparing its outputs with the reference outputs. A hierarchical verifier combines execution and output-schema checks, output comparison, and anti-shortcut checks. Feedback on public tests supports iterative revision. The framework supports additional workflows and input configurations without requiring a separate reference solution for each task. We instantiate the framework as Software Workflows Reconstruction (SWR), which contains 3,000 tasks from 500 workflows in 46 software families across six domains, including physical and engineering simulation, life and molecular sciences, and earth and space sciences. Qwen3.8-Max achieves 22.8\% Pass@1 and 27.9\% Pass@3. We sample verified trajectories to construct 3,000 training trajectories. Supervised fine-tuning of Qwen3.8-27B improves mean Terminal-Bench 2 performance from 47.94\% to 53.56\% across three seeds and yields higher mean scores on all four evaluations than training on any of the four control corpora under the same token budget. These results support using scientific software to collect verified reconstruction trajectories for terminal-agent training.
}

\headercontent{%
\refstepcounter{figure}\label{fig:frontpage-overview}%
\vspace{1.5mm}
\includegraphics[width=\linewidth]{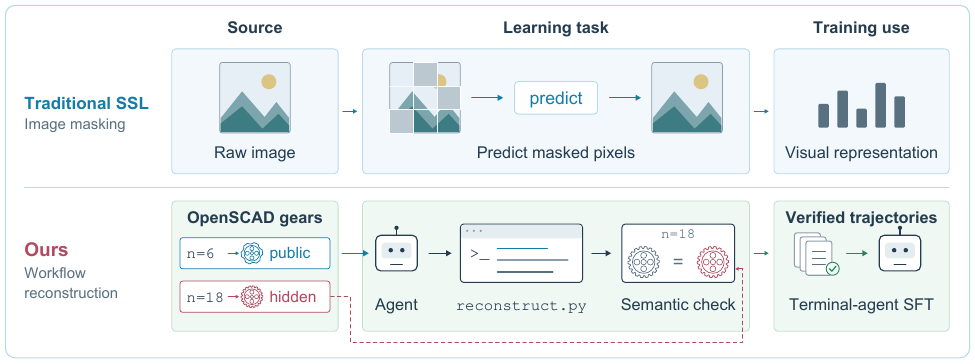}\par
\vspace{0.6mm}

\begin{minipage}{\linewidth}
\small
\justifying
\setlength{\parindent}{0pt}
Figure~\thefigure. Self-supervised learning via workflow reconstruction.
Similar to input masking in conventional self-supervised learning,
our approach masks underlying workflows and verifies their
reconstruction on hidden inputs to generate agent training trajectories.
\par
\end{minipage}
}

\begin{document}
\thispagestyle{firstheader}
\maketitle

\section{Introduction}

Large language model (LLM) agents increasingly operate in terminal environments, where they execute programs, interact with software tools, and solve multi-step tasks \citep{luo2025large,yuan2026terminal}. Their capabilities are commonly evaluated through execution-based benchmarks such as SWE-bench, SWE-Gym, and Terminal-Bench \citep{jimenez2024swebench,pan2025swegym,merrill2026terminalbench}. While existing training environments predominantly focus on software engineering, many scientific and engineering applications rely on specialized computational workflows \citep{li2025aircraft,li2021robot}. Extending terminal-agent capabilities to these domains requires scalable approaches to constructing executable training environments that capture domain-specific workflows and provide reliable verification of agent-generated solutions.

Constructing training environments for scientific and engineering domains presents two fundamental challenges. The first is reliable verification. Scientific workflows often produce heterogeneous outputs, including numerical arrays, meshes, waveforms, biological sequences, and structured tables, whose correctness requires domain-specific evaluation criteria \citep{chen2025scienceagentbench,hu2026scicode}. A reliable verifier must recognize semantically equivalent solutions while rejecting superficially valid but incorrect outputs. The second challenge is scalability. Existing benchmarks often rely on task-specific reference solutions and verification procedures, making environment construction expensive and labor-intensive as task diversity increases \citep{hendrycks2021apps,jain2025livecodebench,zhou2024webarena,shen2026seta,li2026recursive}. These challenges raise a fundamental question: how can we scale the construction of executable training environments without independently engineering reference solutions and verifiers for every task?

To address this problem, we introduce \emph{software-in-the-loop reconstruction}, a self-supervised framework that transforms existing software workflows into executable and verifiable agent tasks. Inspired by masked modeling in self-supervised learning \citep{lin2025prototypes,hondru2025masked}, we extend the reconstruction objective from missing data to underlying computational workflows (Figure~\ref{fig:frontpage-overview}). Specifically, we execute reference workflows under different input configurations to generate public examples and hidden test cases. Given public input--output pairs, an agent reconstructs the workflow's behavior without access to its original implementation, following recent advances in inductive program synthesis \citep{li2025combining,wei2025codearc}. The reconstructed program is evaluated on hidden inputs, where software execution provides verifiable supervision \citep{wei2025codearc,zhao2026absolute}. By reusing reference workflows and semantic comparison rules across input configurations, our framework enables scalable task generation and verified trajectory collection for terminal-agent training across scientific and engineering domains.

\begin{figure}[H]
\centering
\includegraphics[width=\linewidth]{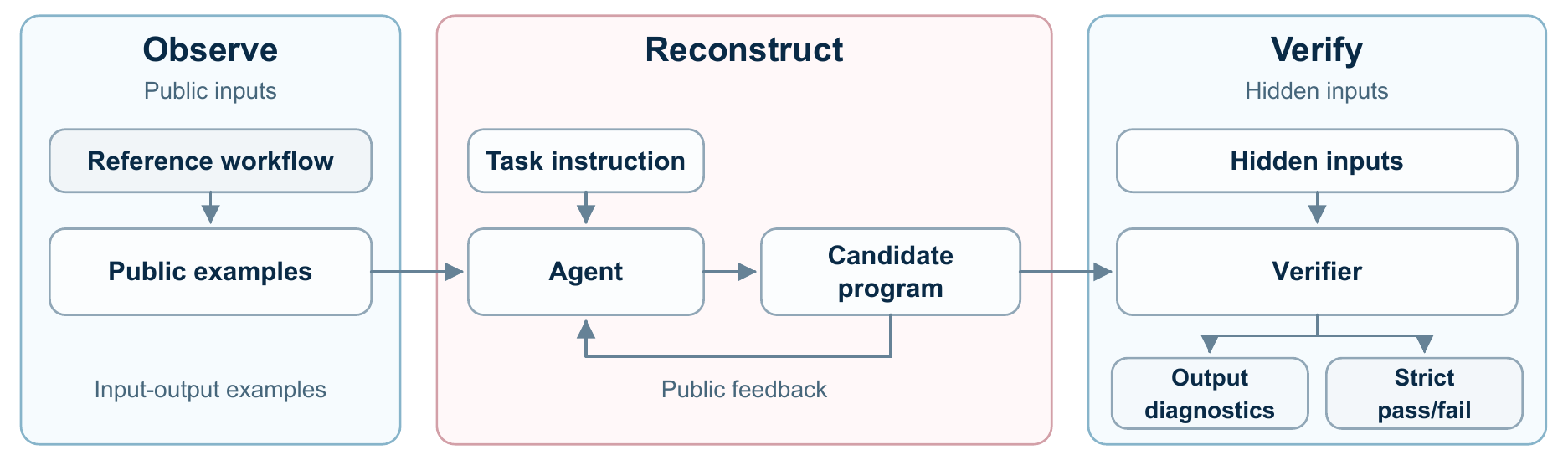}
\caption{Overview of software-in-the-loop reconstruction. The agent reconstructs workflow behavior from public input–output examples, and the verifier evaluates the reconstructed program on hidden inputs by comparing its outputs with reference outputs.}
\label{fig:pipeline}
\end{figure}

Figure~\ref{fig:pipeline} illustrates the overall framework of software-in-the-loop reconstruction. The key idea is to turn existing software workflows into a scalable source of executable and verifiable agent tasks. Rather than designing a reference solution and verifier from scratch for each task, we reuse existing workflows as reference solutions, generate reference outputs through software execution, and share semantic comparison rules across tasks within the same software family. Once a workflow is integrated into the framework, new tasks can be generated automatically by varying input configurations. Following this framework, we construct Software Workflows Reconstruction (SWR), comprising 3,000 tasks derived from 500 workflows across 46 software families and six domains, covering physical and engineering simulation, life and molecular sciences, earth and space sciences, and supporting computational workflows.

Our experiments demonstrate the reliability of task verification, the difficulty of workflow reconstruction, and the effectiveness of the resulting training data. The verifier achieves 98.0\% agreement with human judgments in a blinded review of 200 audit cases conducted after the verifier is frozen. Despite access to public examples and execution feedback, Qwen3.8-Max achieves only 22.8\% Pass@1 and 27.9\% Pass@3 on SWR. More importantly, training on verified reconstruction trajectories substantially improves terminal-agent performance. Using 3,000 sampled training trajectories, supervised fine-tuning (SFT) improves Qwen3.8-27B from 47.94\% to 53.56\% on Terminal-Bench 2.0 (TB2) and Qwen3.5-27B from 41.57\% to 47.57\%. Under the same training token budget, SWR outperforms four existing agent training corpora across all four evaluations: TB2, Terminal-Bench 4.0 (TB4), Long-Horizon Terminal-Bench (LHTB), and SWR100. These results demonstrate the potential of existing scientific software as a scalable source of high-quality training data for terminal agents.

Our main contributions are summarized as follows:
\begin{enumerate}
    \item We introduce software-in-the-loop reconstruction, a self-supervised framework that transforms existing software workflows into executable and verifiable agent tasks. By reusing reference workflows across different input configurations, our framework enables scalable environment generation without manually constructing reference solutions for individual tasks.
    \item We construct Software Workflows Reconstruction (SWR), comprising 3,000 tasks derived from 500 workflows across 46 software families and six domains. We develop a hierarchical verification mechanism to ensure task validity and solution correctness, and systematically evaluate its reliability through controlled stress tests, independent audits, and blinded human evaluation.
    \item We demonstrate the effectiveness of verified reconstruction trajectories for terminal-agent training. Supervised fine-tuning on SWR improves Qwen3.8-27B from 47.94\% to 53.56\% and Qwen3.5-27B from 41.57\% to 47.57\% on Terminal-Bench 2.0. Under an identical budget of 88.48M assistant-loss tokens, SWR also outperforms four existing training corpora across Terminal-Bench 2.0, Terminal-Bench 4.0, Long-Horizon Terminal-Bench, and SWR100 for Qwen3.8-27B.
\end{enumerate}

\section{Related Work}

\paragraph{Executable environments for agents.}
Executable environments provide a foundation for evaluating and training LLM agents through interactions with software tools and execution-based feedback. Early benchmarks evaluate program correctness through test cases \citep{hendrycks2021apps,jain2025livecodebench}, while subsequent work extends evaluation to interactive software environments \citep{yang2023intercode,liu2024agentbench,mialon2024gaia,zhou2024webarena,xie2024osworld}. More recent efforts focus on realistic software engineering and terminal tasks \citep{jimenez2024swebench,yang2024sweagent,pan2025swegym,wang2025openhands,merrill2026terminalbench,li2026long}, alongside scalable approaches to synthesizing executable environments and collecting agent trajectories \citep{raoof2026openthoughts,ivison2026tmax,peng2026litecoder,pi2026data,shen2026seta,li2026recursive}. Complementary research improves training data through feature-guided synthesis and quality-aware distillation \citep{li2026less,li2026beyond,huang2026trust}. In contrast to approaches that synthesize or extend individual tasks, SWR leverages existing software workflows as reusable reference solutions, enabling scalable construction of verifiable training environments across scientific and engineering domains.

\paragraph{Execution-based verification and feedback.}
Reliable verification is fundamental to execution-based evaluation and agent training. The test-oracle problem studies how to determine the correctness of program outputs \citep{barr2015oracle}, with established approaches including differential testing across implementations \citep{mckeeman1998differential} and metamorphic testing across input transformations \citep{chen1998metamorphic}. Recent agent frameworks leverage interaction feedback to support iterative reasoning, implementation, and refinement \citep{yao2023react,shinn2023reflexion,madaan2023selfrefine,wang2024voyager,le2022coderl,wei2025codearc}. However, verifying scientific workflows requires accounting for heterogeneous output structures, numerical tolerances, and domain-specific notions of equivalence. SWR addresses these challenges through hierarchical verification that combines execution checks, output comparison, and anti-shortcut mechanisms. Public examples provide feedback for iterative refinement, while hidden test cases independently evaluate the submitted program and determine which trajectories are retained for training.

\paragraph{Self-supervised learning and workflow reconstruction.}
Self-supervised learning constructs training objectives from the inherent structure of data without requiring manual annotations. Recent advances in masked modeling learn transferable representations by reconstructing withheld information from partial observations \citep{lin2025prototypes,hondru2025masked}. Similar reconstruction principles have been explored in programming by example and system identification, where computational behavior is inferred from input--output observations \citep{gulwani2011string,gulwani2017program,ljung1999system,wei2025codearc,li2025combining}. While conventional self-supervised learning primarily focuses on reconstructing data or learning representations, SWR extends this paradigm to reconstructing computational workflows. By using existing software to generate reference outputs and verify reconstructed programs on hidden inputs, SWR establishes a self-supervised mechanism for scaling executable training environments and collecting verified agent trajectories.

\section{Software-in-the-Loop Reconstruction}

\subsection{Task Definition}

The goal of workflow reconstruction is to reproduce the input--output behavior of an existing software workflow from a task instruction, an input schema, and a set of public input--output examples. Given these observations, the agent develops a candidate program $\gtheta$ within the provided execution environment, using the available tools and software packages. The agent has no access to the reference workflow implementation, hidden inputs, hidden reference outputs, or private verifier state. Public examples provide execution feedback for iterative refinement, while hidden test cases evaluate whether the reconstructed program reproduces the reference workflow's behavior beyond the observed examples.

Formally, we represent a software workflow $d$ as an executable function $f_d$, where $s$ denotes an input configuration and $f_d(s)$ denotes its corresponding output. Each reconstruction task is defined as
\begin{equation}
\begin{aligned}
T &= (d,f_d,\opub,\spub,\shid,\judge,\instr), \\
\spub\cap\shid &= \emptyset,
\qquad
\opub = \{(s,f_d(s)):s\in\spub\}.
\end{aligned}
\label{eq:task-definition}
\end{equation}
Here, $\spub$ and $\shid$ denote the disjoint sets of public and hidden input configurations, respectively, $\opub$ denotes the public input--output examples, $\instr$ specifies the task instruction, and $\judge$ denotes the semantic comparator. The reconstruction objective is to produce a candidate program $\gtheta$ whose outputs agree with those of $f_d$ under $\judge$ for previously unseen inputs.

Figure~\ref{fig:snapfit-case} illustrates this objective through an OpenSCAD snap-fit design task. Given a public example with a box height of 16\,mm, the agent must reconstruct a program that generates the corresponding geometry for a hidden configuration with a height of 21\,mm, while preserving the body and lid components. This requires the agent to capture the underlying relationship between input parameters and workflow outputs rather than simply reproduce the observed examples.

\subsection{Task Generation from Workflows}
\label{sec:task-generation}

Given a reference workflow $f_d$ and its typed input space $\mathcal{S}_d$, we generate multiple reconstruction tasks by sampling different input configurations. For the $k$-th task variant, we sample disjoint public and hidden input sets $\spub^{(k)}, \shid^{(k)} \subset \mathcal{S}_d$ and execute $f_d$ to obtain the corresponding reference outputs:
\begin{equation}
\mathcal{O}_x^{(k)}
=
\left\{(s,f_d(s)):s\in S_x^{(k)}\right\},
\qquad
x\in\{\mathrm{pub},\mathrm{hid}\}.
\label{eq:scenario-generation}
\end{equation}
For example, varying the gear tooth count of an OpenSCAD workflow generates distinct geometric outputs without modifying the reference implementation. In SWR, we instantiate $K=6$ task variants per workflow. Within each variant, input configurations are sampled without replacement using deterministic seeds. We audit scenario overlap across variants after task generation (Appendix~\ref{app:dataset-audit}).

To determine whether a generated task is valid, we define three workflow-level validation functions: $D(T)$ for execution reproducibility, $C(T)$ for output schema consistency, and $I(T)$ for input sensitivity. Each function returns a binary value indicating whether the corresponding criterion is satisfied. Let $g_d^\star$ denote the reference workflow adapted to the reconstruction interface, and let $\mathcal{B}_{\mathrm{pub}}$ denote the set of three predefined public-output replay strategies. A valid task must admit the reference implementation as a correct solution while rejecting all replay strategies. Accordingly, we define the task acceptance criterion as
\begin{equation}
\begin{aligned}
V(T) ={}& D(T)\cdot C(T)\cdot I(T) \\
&\cdot \operatorname{Pass}(T,g_d^\star)
\cdot \prod_{b\in\mathcal{B}_{\mathrm{pub}}}
\left(1-\operatorname{Pass}(T,b)\right).
\end{aligned}
\label{eq:task-validity}
\end{equation}
Here, $\operatorname{Pass}(\cdot,\cdot)$ denotes the strict success criterion defined in Section~\ref{sec:verifier}. Only tasks satisfying $V(T)=1$ are retained. Given a collection of reference workflows $\mathcal{D}$, the resulting reconstruction dataset is defined as
\begin{equation}
\mathcal{D}_{\mathrm{SWR}}
=
\left\{
T_{d,k}:d\in\mathcal{D},\;
1\leq k\leq K,\;
V(T_{d,k})=1
\right\}.
\label{eq:swr-dataset}
\end{equation}
This formulation separates workflow integration from task instantiation: additional tasks are generated by sampling new input configurations while retaining the reference implementation and comparison rules. Appendix~\ref{app:specification} provides the complete generation algorithm, and Appendix~\ref{app:dataset-audit} analyzes the diversity and validity of the resulting tasks.

\subsection{Hierarchical Verification}
\label{sec:verifier}

Given a reconstruction task $T$ and a candidate program $\gtheta$, we evaluate the program through its outputs on a set of input configurations. Let $E(\gtheta,s)\in\{0,1\}$ indicate whether $\gtheta$ executes successfully on input $s$ and produces outputs conforming to the required schema. For valid outputs, we define an output comparator $\judge$ that measures the agreement between $\gtheta(s)$ and $f_d(s)$. The verification score for an individual input configuration is given by
\begin{equation}
v(T,\gtheta,s)=
\begin{cases}
\judge\!\left(\gtheta(s),f_d(s)\right), & E(\gtheta,s)=1,\\
0, & \text{otherwise},
\end{cases}
\label{eq:scenario-verification}
\end{equation}
where $\judge(\cdot,\cdot)\in[0,1]$ is defined according to the output properties of the corresponding software family. Specifically, categorical values are compared by exact matching, numerical values by absolute and relative tolerances, and geometric outputs by domain-specific comparison functions. The execution interface is shared across workflows, whereas the comparison rules are shared across tasks within each software family.

Based on ~\eqref{eq:scenario-verification}, we define the reconstruction score over the public and hidden input sets as
\begin{equation}
r_x(T,\gtheta)
=
\frac{1}{|S_x|}
\sum_{s\in S_x}
v(T,\gtheta,s),
\qquad
x\in\{\mathrm{pub},\mathrm{hid}\}.
\label{eq:semantic-scores}
\end{equation}
During agent interaction, $\rpub$ provides feedback for iterative program refinement. Upon termination of the rollout, we fix $\gtheta$ and compute $\rhid$ on the hidden inputs. In addition to output agreement, let $A(T,\gtheta)\in\{0,1\}$ indicate whether the candidate program satisfies all applicable task-specific anti-shortcut constraints. We define the strict success criterion as
\begin{equation}
\operatorname{Pass}(T,\gtheta)
=
\mathbf{1}\!\left[\rhid=1\right]
\cdot A(T,\gtheta).
\label{eq:strict-success}
\end{equation}
Since $v(T,\gtheta,s)\in[0,1]$, the condition $\rhid=1$ requires successful execution, schema compliance, and complete output agreement for every hidden input. Partial agreement is used only for diagnostic analysis. The comparison rules and verifier implementation are provided in Appendix~\ref{app:specification}, with reliability analyses in Appendix~\ref{app:verifier-stress}.



\begin{figure}[t]
\centering
\includegraphics[width=\linewidth]{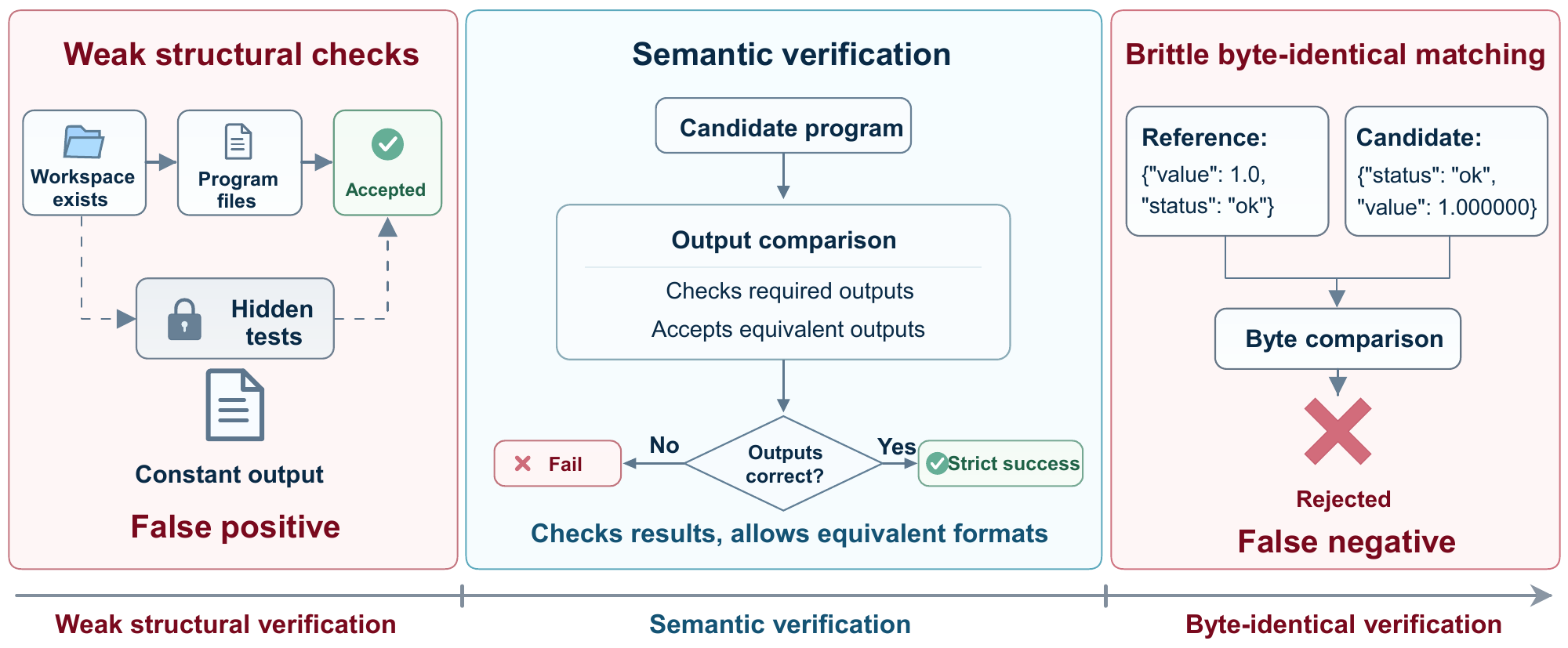}
\caption{Comparison of verification approaches. Our verifier rejects incorrect outputs while accepting equivalent results, addressing the limitations of structural and byte-identical checks.}
\label{fig:verifier}
\end{figure}

\subsection{Reconstruction SFT}
\label{sec:reconstructionsft}

Given the reconstruction dataset $\mathcal{D}_{\mathrm{SWR}}$, we collect agent trajectories under the reconstruction protocol defined above. Let $\tau_i=((h_{i,t},a_{i,t}))_{t=1}^{L_i}$ denote a trajectory collected for task $T_i$, where $L_i$ is the number of agent steps, $h_{i,t}$ denotes the interaction history preceding step $t$, and $a_{i,t}$ denotes the corresponding agent action. Each trajectory yields a final candidate program $g_{\theta_i}$, which is evaluated according to the strict success criterion in ~\eqref{eq:strict-success}. We define the set of verified trajectories as
\begin{equation}
\mathcal{D}_{\mathrm{SFT}}
=
\left\{
\tau_i
\mid
T_i\in\mathcal{D}_{\mathrm{SWR}},
\operatorname{Pass}(T_i,g_{\theta_i})=1
\right\}.
\label{eq:verified-sft}
\end{equation}
Unlike trajectory selection based solely on public execution feedback, this criterion requires the final candidate program to reproduce the reference workflow's behavior on hidden inputs. Each retained trajectory provides supervision for the reconstruction process, including example inspection, program implementation, tool interaction, and iterative refinement.

We construct the training corpus $\mathcal{B}_{\mathrm{SFT}}$ by sampling trajectories from $\mathcal{D}_{\mathrm{SFT}}$ with replacement. Let $p_{\phi}$ denote the student model parameterized by $\phi$. The supervised fine-tuning objective is defined as the negative log-likelihood of the recorded agent actions conditioned on their preceding interaction histories:
\begin{equation}
\mathcal{L}_{\mathrm{SFT}}(\phi)
=
-\sum_{\tau_i\in\mathcal{B}_{\mathrm{SFT}}}
\sum_{t=1}^{L_i}
\log p_{\phi}(a_{i,t}\mid h_{i,t}),
\label{eq:sft-objective}
\end{equation}
where the loss is computed only over assistant-generated tokens, and task instructions and tool feedback are retained as conditioning context. Through this objective, the student model learns from verified workflow reconstruction trajectories without requiring manually annotated solution trajectories.

\section{Experiments}
\label{sec:experiments}

\subsection{Experimental Setup}
\label{sec:experimental-setup}

\paragraph{Evaluation benchmarks.}
We evaluate our framework on both workflow reconstruction and downstream terminal-agent tasks. For workflow reconstruction, we evaluate agents on the full SWR dataset of 3,000 tasks and construct SWR100, a challenging subset comprising 100 tasks from distinct workflows that Qwen3.8-Max fails to solve in three attempts. To assess the transferability of reconstruction training, we evaluate the fine-tuned models on SWR100 and three external benchmarks: Terminal-Bench 2.0 (TB2), Terminal-Bench 4.0 (TB4), and Long-Horizon Terminal-Bench (LHTB) \citep{merrill2026terminalbench,li2026long}. Additional experiments on software families absent from the SFT corpus are presented in Appendix~\ref{app:controlled-studies}.

\paragraph{Agent evaluation.}
We evaluate Qwen3.8-Max-0902 on the full SWR dataset using the Terminus-2 harness, with three attempts per task for performance evaluation and trajectory collection. All three attempts have a wall-clock budget of 60 minutes per task, with a maximum of 100 turns in the first attempt. To further assess reconstruction difficulty, we compare eight API-served models on SWR100: GPT-5.6 Sol, Kimi-K3, Grok 4.6, DeepSeek-V4-Pro, GLM-5.3, MiniMax-M3, Qwen3.8-Max, and Doubao Seed 2.1. Each model is evaluated under a common budget of 500 turns and three hours per task.

\paragraph{Evaluation metrics.}
For SWR, we report Pass@1 and Pass@3 based on the strict success criterion defined in ~\eqref{eq:strict-success}. Pass@1 measures the proportion of tasks solved on the first attempt, while Pass@3 measures the proportion solved in at least one of three attempts. Missing or unscored attempts are treated as failures. For downstream evaluations, we report Pass@1 on TB2, TB4, and SWR100, and the average dense reward on LHTB.

\paragraph{Training setup and baselines.}
We conduct supervised fine-tuning on Qwen3.8-27B and Qwen3.5-27B using verified reconstruction trajectories selected according to ~\eqref{eq:verified-sft}. Trajectories associated with SWR100 are excluded before corpus construction. To evaluate the effect of training scale, we construct corpora containing 750, 1,500, and 3,000 sampled trajectories, allowing repeated sampling. We compare SWR with four existing terminal-agent training corpora: OpenThoughts-Agent-v1 \citep{raoof2026openthoughts}, TMax \citep{ivison2026tmax}, LiteCoder-Terminal \citep{peng2026litecoder}, and Nemotron-Terminal \citep{pi2026data}. For the corpus comparison on Qwen3.8-27B, we fix the training budget at 88.48M assistant-loss tokens across all methods. We report the mean and standard deviation over three independent training and evaluation runs with seeds 42, 43, and 44, using the same evaluation seeds for the base models. Details of trajectory preprocessing and corpus construction are provided in Appendix~\ref{app:sft-corpus}.

\begin{figure*}[t]
\centering
\begin{subfigure}[t]{0.33\textwidth}
\centering
\includegraphics[width=\linewidth]{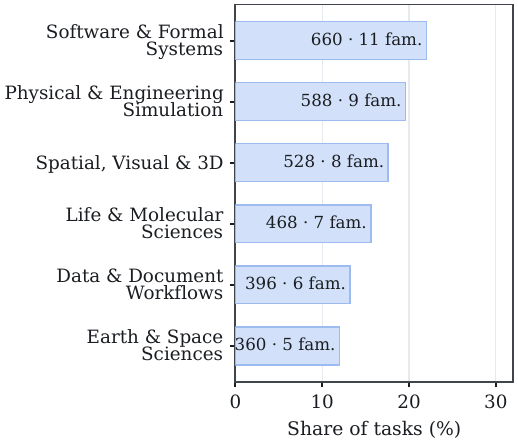}
\caption{Distribution of tasks and software families across domains.}
\label{fig:domain-coverage}
\end{subfigure}\hfill
\begin{subfigure}[t]{0.33\textwidth}
\centering
\includegraphics[width=\linewidth]{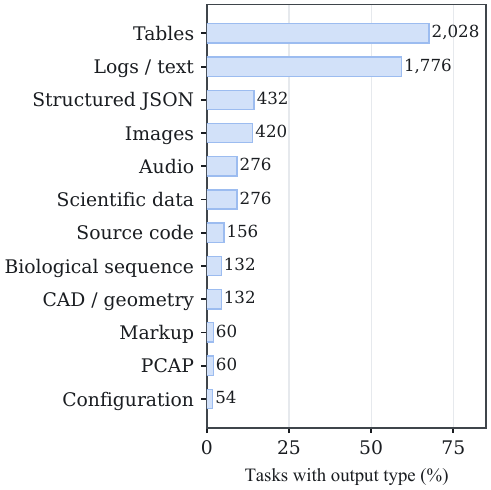}
\caption{Distribution of workflow output types.}
\label{fig:artifact-modalities}
\end{subfigure}\hfill
\begin{subfigure}[t]{0.33\textwidth}
\centering
\includegraphics[width=\linewidth]{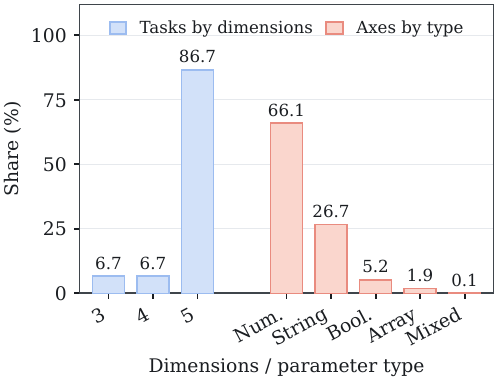}
\caption{Distribution of input dimensions and parameter types.}
\label{fig:parameter-complexity}
\end{subfigure}
\caption{Dataset composition of SWR. Statistics of domain coverage, workflow output types, and input parameter configurations across 3,000 reconstruction tasks.}
\label{fig:dataset-composition}
\end{figure*}

\subsection{Task Coverage and Validation}
\label{sec:dataset-statistics}

\paragraph{Dataset composition.}
As summarized in Table~\ref{tab:dataset}, SWR comprises 3,000 reconstruction tasks derived from 500 reference workflows across 46 software families and six domains. Figure~\ref{fig:dataset-composition} presents the distributions of software domains, output types, and input parameters. The tasks cover diverse computational workflows, ranging from physical simulation and molecular analysis to spatial computing and formal systems, with no individual software family accounting for more than 2.4\% of the dataset. Across all tasks, SWR contains 15,600 public and 8,400 hidden input configurations. The diversity of output types and input parameters reflects the heterogeneous computational requirements of these domains. Appendix~\ref{app:dataset-audit} provides a detailed breakdown of the dataset.

\paragraph{Construction efficiency.}
The construction of SWR requires approximately 336 person-hours from two contributors, primarily for software-family integration, reference workflow implementation, and input space specification. Approximately 20\% of candidate workflows are rejected by the automated validation procedure, resulting in 500 retained workflows. Each retained workflow generates six task variants without additional manual task authoring. By separating workflow integration from task instantiation, the framework amortizes the initial engineering effort across multiple reconstruction tasks. Appendix~\ref{app:specification} details the construction procedure and the scope of the reported effort.

\begin{table}[H]
\caption{Dataset statistics of SWR across six computational domains.}
\label{tab:dataset}
\centering
\small
\setlength{\tabcolsep}{8pt}
\renewcommand{\arraystretch}{1.15}
\begin{tabular}{lrrrr}
\toprule
\textbf{Domain} & \textbf{Families} & \textbf{Workflows} & \textbf{Tasks} & \textbf{Share (\%)} \\
\midrule
Software \& Formal Systems          & 11 & 110 & 660 & 22.0 \\
Physical \& Engineering Simulation  & 9  & 98  & 588 & 19.6 \\
Spatial, Visual \& 3D               & 8  & 88  & 528 & 17.6 \\
Life \& Molecular Sciences          & 7  & 78  & 468 & 15.6 \\
Data \& Document Workflows          & 6  & 66  & 396 & 13.2 \\
Earth \& Space Sciences             & 5  & 60  & 360 & 12.0 \\
\midrule
\textbf{Total} & \textbf{46} & \textbf{500} & \textbf{3,000} & \textbf{100.0} \\
\bottomrule
\end{tabular}
\end{table}

\paragraph{Input--output dependency.}
We analyze the quality of generated tasks by measuring the variation in reference outputs across input configurations. Specifically, we conduct two experiments: (i) pairwise comparison of reference outputs under different input configurations, and (ii) individual parameter perturbation while keeping the remaining parameters fixed.

\begin{figure}[t]
\centering
\includegraphics[width=0.78\linewidth]{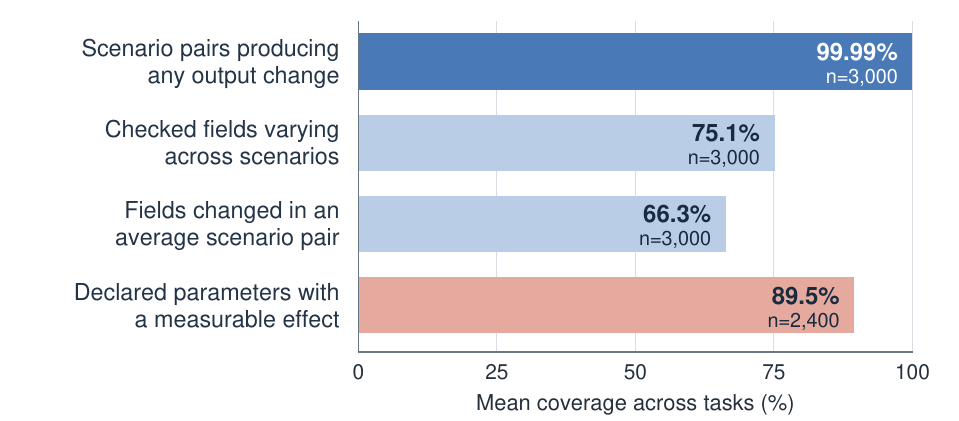}
\caption{Output variation across input configurations and sensitivity to individual input parameters in SWR. The scenario-pair sensitivity is the mean across tasks of the fraction of evaluated scenario pairs that produce different reference outputs, excluding digest-like fields. Scenario-pair sensitivity is reported to two decimal places, and the other percentages to one decimal place.}
\label{fig:scenario-sensitivity}
\end{figure}

\textbf{SWR tasks exhibit substantial output variation across input configurations.} As shown in Figure~\ref{fig:scenario-sensitivity}, all 3,000 tasks exhibit output variation, with 2,984 tasks producing distinct outputs for every pair of input configurations and the remaining 16 exhibiting output changes for 35 of 36 pairs. On average, 75.1\% of checked output fields vary across configurations, and 66.3\% differ between two configurations. Furthermore, across 2,400 tasks evaluated through individual parameter perturbations, 89.5\% of input parameters affect the reference outputs. These results indicate that SWR provides diverse and input-sensitive test cases for evaluating workflow reconstruction, rather than generating redundant task instances from the same workflow.

\paragraph{Shortcut evaluation.}
We evaluate three public-output replay baselines: (i) \emph{most-frequent replay}, which returns the most frequent public output; (ii) \emph{nearest-input replay}, which retrieves the public output associated with the nearest input configuration; and (iii) \emph{oracle-selected replay}, which selects the public output that maximizes agreement with the hidden reference output.

\textbf{Public-output replay achieves limited agreement with hidden reference outputs.} The three replay baselines achieve output-field agreement of 34.3\%, 38.0\%, and 44.3\%, respectively. Even oracle-selected replay, which selects public outputs based on their agreement with hidden reference outputs, reproduces fewer than half of the evaluated output fields on average. All three baselines achieve a strict success rate of 0.0\%, consistent with the replay rejection criterion in \eqref{eq:task-validity}. These results characterize the extent to which hidden outputs can be approximated through direct reuse of public observations. Additional results are provided in Appendix~\ref{app:dataset-audit}.

\subsection{Verifier Reliability}
\label{sec:verifier-reliability}

We evaluate verifier reliability through controlled stress tests, post-calibration audits, and blinded human evaluation. During development, we construct 35,873 test cases covering numerical tolerance boundaries, structured output errors, wrong-scenario outputs, and geometric perturbations. Following calibration, we freeze the verifier and construct an independent audit set of 1,200 cases from 100 tasks covering all 46 software families. None of these audit cases is used for verifier calibration or threshold adjustment.

\textbf{The verifier achieves high agreement with both predefined and human-adjudicated labels.} On the independent audit set of 1,200 cases, the verifier achieves 99.25\% agreement with the predefined validity labels. To further assess its reliability, we conduct a blinded human evaluation of 200 audit cases, comprising 100 valid and 100 invalid outputs after adjudication. The verifier achieves 98.0\% agreement with human judgments, with a false acceptance rate and a false rejection rate of 2.0\% each. These results provide empirical evidence for the reliability of output verification across diverse software families. The remaining disagreements primarily arise from numerical perturbations near tolerance boundaries and alternative mesh tessellations that preserve the underlying surface geometry. Appendix~\ref{app:verifier-stress} presents the evaluation protocol and detailed error analysis, while Section~\ref{sec:sft-results} examines the effect of task-validation components on downstream training.

\subsection{Agent Performance}
\label{sec:rollout-results}

\paragraph{Reconstruction difficulty and failure analysis.}
Table~\ref{tab:rollouts} reports the performance of Qwen3.8-Max on the full SWR dataset. Despite access to public input--output examples and iterative execution feedback, Qwen3.8-Max achieves only 22.8\% Pass@1 and 27.9\% Pass@3. Increasing the number of attempts from one to three improves the success rate by 5.1 percentage points, indicating that a substantial proportion of reconstruction tasks remain unsolved even with repeated attempts.

\textbf{The primary challenge is reproducing workflow behavior rather than generating executable programs.} Among the executable candidate programs in 2,781 evaluated workspaces, 1,591 return identical outputs for different hidden inputs, and 75 directly replay public outputs. Moreover, successful reconstructions require a median of 58 interaction turns, compared with 85 turns for model-attributable failures. These observations suggest that additional interaction alone is insufficient to resolve reconstruction failures, particularly when agents fail to capture the relationship between input parameters and reference outputs. Appendix~\ref{app:frozen-diagnostics} provides detailed failure statistics.

\begin{table}[H]
\caption{Reconstruction performance and failure analysis of Qwen3.8-Max on SWR.}
\label{tab:rollouts}
\centering
\small
\setlength{\tabcolsep}{12pt}
\renewcommand{\arraystretch}{1.12}
\begin{tabular}{llr}
\toprule
\textbf{Category} & \textbf{Metric} & \textbf{Result} \\
\midrule
\multirow{3}{*}{Reconstruction Performance}
& Pass@1 & 22.8\% \\
& Pass@3 & 27.9\% \\
& Improvement (pp) & +5.1 \\
\midrule
\multirow{4}{*}{Failure Analysis}
& Evaluated workspaces & 2,781 \\
& Identical-output cases & 1,591 \\
& Public-output replay cases & 75 \\
& Failed-run output agreement & 75.4\% \\
\midrule
\multirow{2}{*}{Interaction Cost}
& Median turns (success) & 58 \\
& Median turns (failure) & 85 \\
\bottomrule
\end{tabular}
\end{table}

\begin{figure*}[t]
\centering
\begin{subfigure}[t]{0.40\textwidth}
\centering
\includegraphics[width=\linewidth]{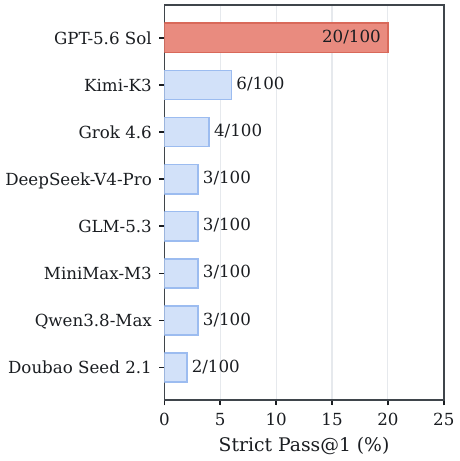}
\caption{Overall reconstruction performance.}
\label{fig:swr100-overall}
\end{subfigure}\hfill
\begin{subfigure}[t]{0.58\textwidth}
\centering
\includegraphics[width=\linewidth]{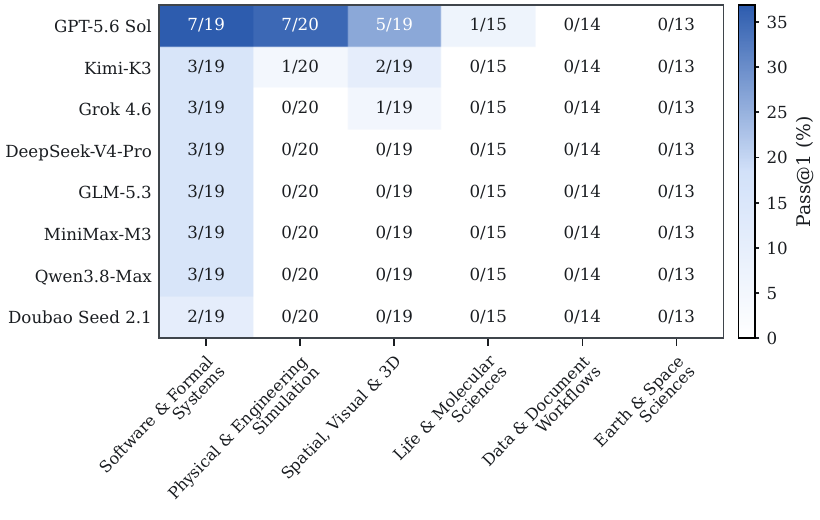}
\caption{Reconstruction performance across domains.}
\label{fig:swr100-domain}
\end{subfigure}
\caption{Performance comparison of eight foundation models on SWR100, including overall Pass@1 and domain-level success rates.}
\label{fig:swr100-model-domain}
\end{figure*}

\paragraph{Comparison across foundation models.}
We evaluate eight API-served models on SWR100 under identical interaction limits. As shown in Figure~\ref{fig:swr100-model-domain}, GPT-5.6 Sol achieves the highest Pass@1 of 20.0\%, outperforming the second-best model, Kimi-K3, by 14.0 percentage points. All seven other models achieve success rates between 2.0\% and 6.0\%.

\textbf{Reconstruction performance varies substantially across scientific domains.} Among the 44 successful reconstructions across all eight models, 43 occur in Software \& Formal Systems, Physical \& Engineering Simulation, and Spatial, Visual \& 3D. Only one successful reconstruction occurs in Life \& Molecular Sciences, while no model solves any task in Data \& Document Workflows or Earth \& Space Sciences. These results reveal substantial limitations of current foundation models on specialized computational workflows within SWR100.

\begin{figure}[H]
\centering
\includegraphics[width=0.96\linewidth]{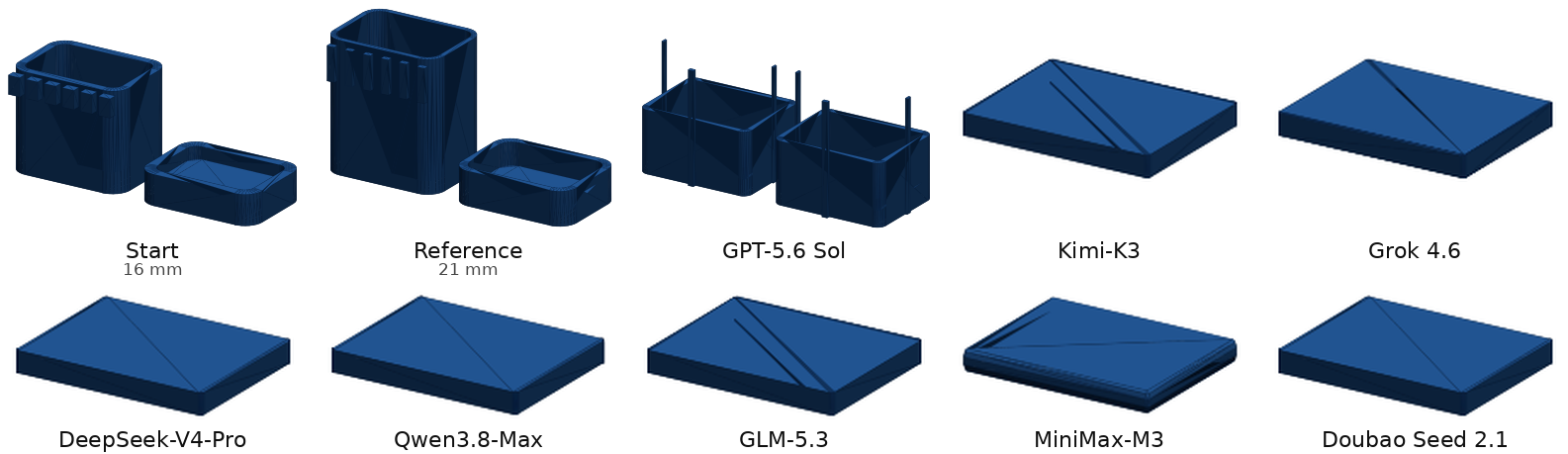}
\caption{Qualitative comparison of eight models on an OpenSCAD snap-fit reconstruction task. GPT-5.6 Sol preserves the required geometry under the hidden input configuration, while the remaining models produce incorrect geometric structures.}
\label{fig:snapfit-case}
\end{figure}

\paragraph{Qualitative analysis.}
Figure~\ref{fig:snapfit-case} presents the reconstructed outputs of eight models on an OpenSCAD snap-fit design task. Given a public example with a box height of 16\,mm, the models are evaluated on a hidden configuration with a height of 21\,mm. GPT-5.6 Sol successfully preserves the geometric structures of the box body and lid while adapting to the new height. In contrast, the remaining seven models produce plate-like geometries whose heights correspond to the wall thickness rather than the specified box height. Although these outputs are valid mesh files, they fail to reproduce the reference workflow's behavior. This example further illustrates the limitations of structural verification and the importance of evaluating output correctness under different input configurations.

\subsection{Reconstruction SFT}
\label{sec:sft-results}

We investigate whether verified workflow reconstruction trajectories can improve terminal-agent capabilities beyond the reconstruction tasks used for training. We first examine the effectiveness of reconstruction SFT and compare SWR with existing training corpora under matched token budgets. We then control for differences in trajectory collection and examine the contribution of task validation. Finally, we evaluate the transferability of reconstruction training across student models and software families.

\paragraph{Effectiveness of reconstruction SFT.}
We first examine whether workflow reconstruction provides effective supervision for terminal-agent training. Starting from Qwen3.8-27B, we construct three training corpora containing 750, 1,500, and 3,000 entries by sampling from 1,422 unique verified trajectories. Table~\ref{tab:sft-results} reports the performance across four evaluations.

\textbf{Reconstruction SFT improves performance across both in-domain and external benchmarks.} With 3,000 training entries, the model improves from 47.94\% to 53.56\% on TB2, from 0.51\% to 3.54\% on TB4, and from 20.67\% to 27.67\% on LHTB. Performance on SWR100 also increases from 1.33\% to 3.33\%. Moreover, the mean performance on each benchmark is non-decreasing as the number of training entries increases. Since the training corpus consists exclusively of workflow reconstruction trajectories, the improvements on TB2, TB4, and LHTB indicate that reconstruction training benefits general terminal-agent capabilities beyond the original training tasks. Appendix~\ref{app:controlled-studies} further examines the effects of trajectory diversity and repeated sampling.

\paragraph{Comparison with existing training corpora.}
Having established the effectiveness of reconstruction SFT, we examine whether reconstruction trajectories provide more effective supervision than existing terminal-agent training data. We compare SWR with four representative training corpora: OpenThoughts-Agent-v1 \citep{raoof2026openthoughts}, TMax \citep{ivison2026tmax}, LiteCoder-Terminal \citep{peng2026litecoder}, and Nemotron-Terminal \citep{pi2026data}. For a fair comparison, we fine-tune the same Qwen3.8-27B base model using 88.48M assistant-loss tokens for each corpus.

\begin{table}[H]
\caption{Effect of reconstruction SFT on Qwen3.8-27B under different training data budgets. Results are reported as mean and standard deviation (\%) over three independent runs.}
\label{tab:sft-results}
\centering
\small
\setlength{\tabcolsep}{5pt}
\begin{tabular}{lrrrr}
\toprule
Training entries & TB2 & TB4 & LHTB & SWR100 \\
\midrule
Base & $47.94 \pm 1.30$ & $0.51 \pm 0.87$ & $20.67 \pm 1.15$ & $1.33 \pm 0.58$ \\
SFT-750 & $50.19 \pm 0.65$ & $1.01 \pm 0.87$ & $22.33 \pm 0.58$ & $1.33 \pm 0.58$ \\
SFT-1500 & $50.56 \pm 1.12$ & $2.02 \pm 0.87$ & $24.33 \pm 1.53$ & $2.67 \pm 0.58$ \\
SFT-3000 & $\mathbf{53.56 \pm 0.65}$ & $\mathbf{3.54 \pm 0.87}$ & $\mathbf{27.67 \pm 0.58}$ & $\mathbf{3.33 \pm 0.58}$ \\
\bottomrule
\end{tabular}
\end{table}

\textbf{SWR achieves the highest mean performance across all four evaluations under the matched-token setting.} As shown in Table~\ref{tab:sft-corpus-controls}, SWR outperforms the strongest competing corpus on each benchmark by 4.12, 2.53, 4.00, and 1.66 percentage points on TB2, TB4, LHTB, and SWR100, respectively. Notably, although SWR is constructed from workflow reconstruction tasks rather than conventional terminal tasks, it achieves higher performance on three external terminal-agent benchmarks. These results demonstrate the effectiveness of workflow reconstruction as a training paradigm for improving terminal-agent performance across diverse downstream tasks.

\begin{table}[H]
\caption{Comparison with existing terminal-agent training corpora under matched token budgets. All corpora are used to fine-tune Qwen3.8-27B with 88.48M assistant-loss tokens. Results are reported as mean and standard deviation (\%) over three independent runs.}
\label{tab:sft-corpus-controls}
\centering
\small
\setlength{\tabcolsep}{4pt}
\resizebox{0.87\linewidth}{!}{
\begin{tabular}{lrrrrr}
\toprule
Training corpus & Loss tokens & TB2 & TB4 & LHTB & SWR100 \\
\midrule
Base (no SFT) & -- & $47.94 \pm 1.30$ & $0.51 \pm 0.87$ & $20.67 \pm 1.15$ & $1.33 \pm 0.58$ \\
OpenThoughts-Agent-v1 & 88.48M & $49.06 \pm 1.30$ & $0.00 \pm 0.00$ & $22.67 \pm 1.53$ & $1.67 \pm 0.58$ \\
TMax & 88.48M & $49.44 \pm 2.25$ & $1.01 \pm 0.87$ & $20.67 \pm 1.53$ & $1.33 \pm 0.58$ \\
LiteCoder-Terminal & 88.48M & $48.69 \pm 1.72$ & $0.51 \pm 0.87$ & $23.00 \pm 1.00$ & $1.00 \pm 1.00$ \\
Nemotron-Terminal & 88.48M & $47.19 \pm 1.12$ & $0.51 \pm 0.87$ & $23.67 \pm 1.15$ & $1.33 \pm 1.53$ \\
\midrule
SWR (ours) & 88.48M & $\mathbf{53.56 \pm 0.65}$ & $\mathbf{3.54 \pm 0.87}$ & $\mathbf{27.67 \pm 0.58}$ & $\mathbf{3.33 \pm 0.58}$ \\
\bottomrule
\end{tabular}
}
\end{table}

\paragraph{Controlled comparison of training trajectories.}
The preceding comparison does not control for differences in the teacher models and trajectory collection procedures used to construct existing training corpora. To address this limitation, we compare three sources of training data: (i) verified reconstruction trajectories from SWR, (ii) conventional terminal-task trajectories collected from TMax-15K, and (iii) failed reconstruction trajectories from SWR. All trajectories are collected using Qwen3.8-Max-0902 with the same Terminus-2 harness and rollout limits. We further control the student model, training configuration, and assistant-loss-token budget.

\textbf{The advantage of verified reconstruction trajectories persists under controlled collection settings.} Training on verified SWR trajectories achieves 53.56\% on TB2 and 3.54\% on TB4, compared with 49.06\% and 2.02\% for TMax-15K, respectively. Since both corpora are collected using the same teacher and interaction protocol, the improvement cannot be explained solely by differences in teacher capabilities or rollout budgets. In addition, training on failed SWR trajectories achieves only 46.82\% on TB2 and 1.01\% on TB4. The performance gap between verified and failed reconstruction trajectories highlights the importance of selecting successful trajectories through execution-based verification. Appendix~\ref{app:controlled-studies} provides the complete experimental settings and per-seed results.

\paragraph{Ablation of task-validation components.}
While the preceding experiment examines the importance of trajectory verification, task validity is established before trajectory collection. We therefore investigate whether the validation checks introduced in Section~\ref{sec:task-generation} contribute to the quality of generated tasks and their downstream training utility. Specifically, we separately remove output comparison and public-output replay checks during task construction, while retaining the complete verifier for subsequent trajectory selection.

\textbf{Removing task-validation components increases dataset size but reduces task quality and downstream performance.} Without output comparison, the number of accepted tasks increases from 3,000 to 3,348, but 8.0\% of the resulting tasks are identified as invalid. The corresponding TB2 performance decreases from 53.56\% to 48.31\%. Similarly, removing public-output replay checks yields 3,252 tasks, of which 6.8\% can be solved through replay strategies, and reduces TB2 performance to 50.19\%. Although both ablated settings produce larger task sets, neither achieves the performance of the full pipeline. These results support the importance of task validation in constructing effective training environments, although differences in training token budgets may also contribute to the observed performance gaps. Detailed results are provided in Appendix~\ref{app:verifier-stress}.

\paragraph{Transfer across student models and software families.}
Finally, we examine whether the benefits of reconstruction SFT extend beyond the primary student model and the software families represented in the training corpus. To evaluate transfer across student models, we apply reconstruction SFT to Qwen3.5-27B, whose TB2 performance improves from 41.57\% to 47.57\%. The improvement of 6.00 percentage points is comparable to the 5.62-point improvement observed on Qwen3.8-27B, indicating that the training benefits are not specific to a single student version.

To examine transfer beyond the software families represented in the training corpus, we evaluate a subset of 46 SWR100 tasks from 20 software families that contribute no successful trajectories to reconstruction SFT. The mean success rate increases from 1.45\% to 4.35\%, with improvements observed in three domains. Although the absolute success rates remain low, these results provide preliminary evidence that reconstruction training transfers beyond the software families encountered during SFT. Appendix~\ref{app:controlled-studies} reports the subset selection protocol and domain-level results.

\FloatBarrier
\flushbottom

\section{Conclusion}

We propose software-in-the-loop reconstruction, a self-supervised framework that leverages existing software workflows to scale the construction of verifiable agent training environments. Experiments demonstrate that verified reconstruction trajectories improve terminal-agent performance across multiple benchmarks and two student models. For Qwen3.8-27B, reconstruction SFT also outperforms four existing training corpora across all four evaluations under matched token budgets. These findings establish the training value of scientific workflow reconstruction beyond its original application domains, suggesting that diverse computational workflows can provide transferable supervision for terminal agents. Future work will investigate more complex scientific workflows and the integration of reconstruction-based supervision into reinforcement learning.

\bibliography{iclr2027_conference}
\bibliographystyle{iclr2027_conference}

\clearpage
\appendix
\raggedbottom

\section*{Appendix}

The appendix provides additional methodological details, comprehensive dataset analyses, and supplementary experiments to support the main findings. The remaining sections are organized as follows.

\begin{itemize}
\setlength{\itemsep}{0.75em}

\item \textbf{A. Benchmark Specification.}
Section~\ref{app:specification} presents the complete task generation algorithm, including workflow integration, input configuration sampling, reference output generation, and task validation. We further specify the reconstruction interface, verification criteria, evaluation protocol, and training configurations to facilitate reproducibility.

\item \textbf{B. Dataset Composition and Quality Analysis.}
Section~\ref{app:dataset-audit} provides a comprehensive characterization of SWR across software domains, workflow families, input parameters, and output types. Additional analyses examine output variation under input perturbations, resistance to public-output replay, and the diversity of task variants derived from the same reference workflow.

\item \textbf{C. Reconstruction Performance Analysis.}
Section~\ref{app:frozen-diagnostics} presents detailed reconstruction results on the full SWR dataset, with particular attention to discrepancies between program executability and output correctness. We further analyze interaction efficiency, common failure patterns, and the relationship between reconstruction difficulty and interaction length across software families.

\item \textbf{D. Reconstruction SFT Data Construction.}
Section~\ref{app:sft-corpus} details the construction of the reconstruction-only training corpus from verified agent trajectories. We describe trajectory preprocessing, duplicate-event removal, conversation structure repair, sequence-length filtering, and repeated sampling of successful trajectories.

\item \textbf{E. Additional SFT Experiments.}
Section~\ref{app:sft-pilot} presents the complete per-seed results for reconstruction SFT under different training data and token budgets. Additional experiments examine training performance under controlled trajectory collection, the effects of trajectory diversity, repeated sampling, and task-variant count, as well as transfer across student models and software families absent from SFT.

\item \textbf{F. Model Performance on SWR100.}
Section~\ref{app:swr100-results} presents a detailed comparison of eight foundation models on SWR100. We examine overall reconstruction performance, evaluation completeness, and the distribution of successful reconstructions across six computational domains to characterize domain-specific differences in model capabilities.

\item \textbf{G. Interaction Efficiency and Statistical Analysis.}
Section~\ref{app:swr100-diagnostics} examines the computational and interaction costs of workflow reconstruction across different models, including the number of interaction turns and token consumption. We also discuss incomplete evaluations and the statistical limitations of single-attempt model comparisons on SWR100.

\item \textbf{H. Verifier Reliability and Ablation Studies.}
Section~\ref{app:verifier-stress} systematically evaluates verifier reliability through numerical perturbations, structured output modifications, wrong-scenario substitutions, and geometric transformations. Independent post-freeze audits and blinded human evaluations further assess verification accuracy. We also investigate the contributions of individual task-validation components to dataset quality and downstream SFT performance.

\item \textbf{I. Reconstruction Failure Analysis.}
Section~\ref{app:trajectory-case-studies} provides a fine-grained analysis of failure modes across 800 model--task pairs on SWR100. We characterize model-specific failure distributions and co-occurring errors, distinguishing failures in program generation, execution, output schema compliance, numerical correctness, categorical agreement, and geometric reconstruction.

\end{itemize}

\clearpage

\section{Benchmark Specification}
\label{app:specification}

\subsection{Workflow Integration and Task Generation}

\paragraph{Reference Workflows.}
We construct reference workflows from publicly available scientific and engineering software, documentation, APIs, and executable examples, including OpenFOAM tutorial cases, KiCad PCB projects, and MuJoCo Menagerie robot models. For each software family, we implement reusable adapters for workflow execution and output extraction. Each workflow $f_d$ is associated with a typed input space $\mathcal{S}_d$, which specifies the configurable parameters, their admissible ranges, and the corresponding input constraints.

The construction of SWR involves two contributors working for 21 days (336 person-hours), including 15 days for the implementation of family-level adapters and reference workflows, and six days for the specification of input parameters and ranges. Approximately 20\% of candidate workflows are rejected due to execution failures, non-reproducible outputs, insufficient input sensitivity, or susceptibility to public-output replay. The retained 500 workflows are used for automated task generation.

\paragraph{Scenario Generation.}
For each reference workflow $f_d$, we instantiate $K=6$ reconstruction tasks by sampling public and hidden scenarios from $\mathcal{S}_d$. Let $\spub^{(k)}$ and $\shid^{(k)}$ denote the public and hidden scenario sets of the $k$-th task variant, respectively, where $\spub^{(k)}\cap\shid^{(k)}=\emptyset$. The corresponding reference observations are obtained by executing $f_d$ on the sampled inputs:
\begin{equation}
\mathcal{O}_{j}^{(k)}
=
\left\{(s,f_d(s)) \mid s\in\mathcal{S}_{j}^{(k)}\right\},
\quad j\in\{\mathrm{pub},\mathrm{hid}\}.
\end{equation}
The resulting dataset consists of 3,000 reconstruction tasks, with 15,600 public scenarios and 8,400 hidden scenarios, corresponding to 5.2 public and 2.8 hidden scenarios per task on average.

\paragraph{Task Validation.}
Following the task acceptance criterion in ~\eqref{eq:task-validity}, we verify the executability, reproducibility, schema consistency, and input sensitivity of each reference workflow on the sampled scenarios. Specifically, the reference workflow is executed twice on each public and hidden scenario to examine output reproducibility. We further require that variations in the input configurations induce measurable changes in the required outputs.

Let $g_d^\star$ denote the reference workflow adapted to the reconstruction interface. For each task $T_{d,k}$, we require $\operatorname{Pass}(T_{d,k},g_d^\star)=1$. To exclude tasks solvable through direct reuse of public outputs, we additionally consider the following three replay controls:

\begin{itemize}
    \item \textbf{Most-Frequent-Output Replay.} Returns the most frequently occurring output among the public reference observations.

    \item \textbf{Nearest-Input Replay.} Returns the public reference output associated with the input configuration nearest to the hidden input.

    \item \textbf{Oracle-Selected Replay.} Selects the public reference output that achieves the highest agreement with the hidden reference output under the task-specific comparator.
\end{itemize}

A task is retained only if the reference implementation passes the verifier and all three replay controls fail. The oracle-selected replay control has access to hidden reference outputs and is used exclusively during task validation. The complete task generation procedure is presented in Algorithm~\ref{alg:swr-synthesis}.

\subsection{Reconstruction Interface and Verification}

\paragraph{Reconstruction Interface.}
Each reconstruction task is specified by a task instruction $\instr$, a typed input schema, public reference observations $\opub$, and an execution environment containing the available software dependencies. The reference workflow $f_d$, hidden scenarios $\shid$, hidden reference outputs, and private verifier state are withheld from the agent. The agent constructs a candidate program $\gtheta$ that accepts a JSON-encoded scenario and returns a JSON object containing the required output values or references to generated artifacts. The reconstruction objective is defined with respect to the input--output behavior of $f_d$, without imposing implementation equivalence between $\gtheta$ and $f_d$. Agents may invoke available software libraries and executable tools subject to the constraints specified in the task instruction.

\begin{algorithm}[!t]
\caption{Software-in-the-loop synthesis of reconstruction tasks.}
\label{alg:swr-synthesis}
\begin{algorithmic}[1]
\Require Reference workflows $\mathcal{D}=\{(d,f_d)\}_{d=1}^{500}$, typed input spaces $\{\mathcal{S}_d\}$, number of variants $K=6$
\Ensure Reconstruction dataset $\mathcal{D}_{\mathrm{SWR}}$
\State $\mathcal{D}_{\mathrm{SWR}}\gets\emptyset$
\For{each $(d,f_d)\in\mathcal{D}$}
    \State $g_d^\star\gets\Call{WrapWorkflow}{f_d}$
    \For{$k=1,\ldots,K$}
        \Repeat
            \Statex \textit{// Scenario generation and reference execution}
            \State $(\spub^{(k)},\shid^{(k)})\gets
            \Call{SampleDistinctScenarios}{d,k}$
            \State $valid\gets
            [\spub^{(k)}\cap\shid^{(k)}=\emptyset]$
            \If{$valid$}
                \State $S^{(k)}\gets\spub^{(k)}\cup\shid^{(k)}$
                \State Execute $f_d$ twice on each $s\in S^{(k)}$
                \State $valid\gets
                \Call{Successful}{f_d}\land
                \Call{Deterministic}{f_d}$
                \State $valid\gets valid\land
                \Call{SchemaStable}{f_d}\land
                \Call{InputSensitive}{f_d}$
            \EndIf
            \Statex \textit{// Reconstruction task instantiation}
            \If{$valid$}
                \State $\opub^{(k)}\gets
                \{(s,f_d(s)):s\in\spub^{(k)}\}$
                \State $\mathcal{O}_{\mathrm{hid}}^{(k)}\gets
                \{(s,f_d(s)):s\in\shid^{(k)}\}$
                \State $\instr^{(k)}\gets
                \Call{RenderInstruction}{d,\opub^{(k)}}$
                \State $\judge^{(k)}\gets
                \Call{BuildComparator}{f_d,\mathcal{O}_{\mathrm{hid}}^{(k)}}$
                \State $T_{d,k}\gets
                (d,f_d,\opub^{(k)},\spub^{(k)},
                \shid^{(k)},\judge^{(k)},\instr^{(k)})$
                \State \Call{Materialize}{$T_{d,k}$, environment, public feedback, private verifier}
            \EndIf
            \Statex \textit{// Reference verification and replay rejection}
            \If{$valid$}
                \State $\mathcal{B}_{\mathrm{pub}}\gets
                \Call{BuildReplayControls}{T_{d,k}}$
                \State $valid\gets
                [\operatorname{Pass}(T_{d,k},g_d^\star)=1]$
                \For{each $b\in\mathcal{B}_{\mathrm{pub}}$}
                    \State $valid\gets valid\land
                    [\operatorname{Pass}(T_{d,k},b)=0]$
                \EndFor
            \EndIf
        \Until{$valid$}
        \State $\mathcal{D}_{\mathrm{SWR}}\gets
        \mathcal{D}_{\mathrm{SWR}}\cup\{T_{d,k}\}$
    \EndFor
\EndFor
\State \textbf{assert} $|\mathcal{D}_{\mathrm{SWR}}|=500K=3000$
\State \Return $\mathcal{D}_{\mathrm{SWR}}$
\end{algorithmic}
\end{algorithm}

\paragraph{Public Feedback and Hidden Evaluation.}
During agent interaction, candidate programs are evaluated on public scenarios. The feedback includes execution status, output schema compliance, and agreement with the corresponding reference outputs. Once the agent submits its implementation or reaches the interaction limit, the candidate program is fixed and evaluated on the hidden scenarios. Both public and hidden evaluations use the same execution interface and family-specific comparison rules. Hidden inputs, reference outputs, and evaluation feedback are not available during program refinement.

\paragraph{Output Comparison.}
We implement family-specific comparators to evaluate heterogeneous computational outputs. For 2,940 tasks, the comparator recursively evaluates all required output fields, while the remaining 60 geometry tasks employ domain-specific geometric comparison functions. Categorical values, Boolean values, and integers are evaluated using exact matching. For floating-point outputs, the default absolute and relative tolerances are $10^{-6}$ and $10^{-4}$, respectively. Audio and geospatial outputs employ family-specific relative tolerances, while geometric outputs are evaluated using output-specific comparison criteria. Detailed comparator configurations and controlled perturbation experiments are provided in Appendix~\ref{app:verifier-stress}.

\paragraph{Evaluation Metrics.}
Following Section~\ref{sec:verifier}, we report both binary reconstruction success and continuous diagnostic scores. The binary success indicator follows ~\eqref{eq:strict-success}, requiring agreement with the reference outputs across all hidden scenarios and satisfaction of the applicable task-specific anti-shortcut checks. The continuous score $\rhid$ quantifies partial output agreement and is retained for diagnostic analysis. Reconstructions with incomplete output agreement are counted as failures under the strict success criterion, irrespective of their diagnostic scores.

\subsection{Experimental Setup}

\paragraph{Full-Suite Evaluation.}
We evaluate Qwen3.8-Max on all 3,000 reconstruction tasks using the Terminus-2 harness. Each of the three attempts adopts a 60-minute wall-clock limit per task. The first attempt is limited to 100 interaction turns. We report Pass@1 and Pass@3, where the latter measures the proportion of tasks successfully reconstructed in at least one of three attempts. Infrastructure failures and model-attributable failures are recorded separately for diagnostic analysis. The reported success rates are computed over all scheduled tasks, with missing or unscored attempts counted as failures.

\paragraph{SWR100.}
We construct SWR100 from tasks that Qwen3.8-Max fails to solve in all three attempts. We select 100 tasks from 100 distinct reference workflows while retaining coverage of all 46 software families. Each model--task pair is evaluated once under a maximum of 500 interaction turns and a three-hour wall-clock limit. Since SWR100 is conditioned on teacher failures, its performance estimates are not directly comparable with those obtained on the full SWR dataset under a different interaction budget.

\paragraph{Training and Evaluation Separation.}
All trajectories associated with the 100 SWR100 tasks are excluded from the reconstruction-SFT corpus. This exclusion is performed at the task level rather than the workflow or software-family level. To examine transfer to software families absent from reconstruction SFT, we additionally evaluate a subset of 46 SWR100 tasks covering 20 software families that contribute no successful training trajectories (Appendix~\ref{app:controlled-studies}). We further evaluate downstream terminal-agent performance on TB2, TB4, and LHTB, following the experimental settings in Section~\ref{sec:sft-results}.

\begin{table}[H]
\caption{Experimental protocols for SWR and downstream terminal-agent evaluations. The time limit is specified per attempt. Dashes indicate unspecified settings rather than unlimited interaction budgets.}
\label{tab:evaluation-protocols}
\centering
\small
\setlength{\tabcolsep}{5pt}
\begin{tabular}{lrrrll}
\toprule
Evaluation & Tasks & Runs & Turn Limit & Time Limit & Metric \\
\midrule
SWR & 3,000 & 1 & 100 & 1 hour & Pass@1 \\
SWR & 3,000 & 3 & -- & 1 hour & Pass@3 \\
SWR100 & 100 & 1 & 500 & 3 hours & Pass@1 \\
SFT evaluation & 4 suites & 3 seeds & -- & -- & Pass@1 / Dense Reward \\
\bottomrule
\end{tabular}
\end{table}

\paragraph{Inference Configuration.}
For API-based model evaluation on SWR100, we adopt a temperature of 0.2, top-$p$ of 0.95, and a maximum output length of 8,192 tokens per model invocation. The interaction and wall-clock budgets are specified in Table~\ref{tab:evaluation-protocols}.

\subsection{Training Details}

\paragraph{Model Initialization.}
We initialize Qwen3.8-27B from the official Hugging Face checkpoint \texttt{Qwen/Qwen3.8-27B}. The pretrained weights are converted to the Megatron \texttt{torch\_dist} checkpoint format with the corresponding parameter layout.

\paragraph{Distributed Training.}
We perform full-parameter SFT using SLIME on eight compute nodes, each equipped with eight NVIDIA GPUs (64 GPUs in total). We employ tensor parallelism of 4, pipeline parallelism of 2, context parallelism of 2, and sequence parallelism, with a global batch size of 128. The maximum context length is set to 262,145 tokens, and dynamic packing is limited to 49,152 tokens per GPU.

\paragraph{Optimization.}
We optimize the SFT objective defined in ~\eqref{eq:sft-objective} using sample-level \texttt{sft\_loss} with \texttt{loss\_mask\_type=qwen3\_5}, restricting the training loss to assistant tokens. We use the Adam optimizer with $(\beta_1,\beta_2)=(0.9,0.98)$, a cosine learning rate schedule from $1\times10^{-5}$ to $1\times10^{-6}$, a warmup ratio of 0.05, and zero weight decay. Optimizer and random number generator states are neither restored nor saved.

\paragraph{Training and Evaluation Configuration.}
We use five training steps for reconstruction SFT. For downstream evaluation, we use the same evaluation settings for the base and fine-tuned models and report the mean and standard deviation over three runs with seeds 42, 43, and 44.

\begin{table}[H]
\caption{Training configurations for reconstruction SFT on Qwen3.8-27B.}
\label{tab:sft-hyperparameters}
\centering
\small
\setlength{\tabcolsep}{10pt}
\begin{tabular}{ll}
\toprule
Configuration & Value \\
\midrule
\multicolumn{2}{l}{\textit{Model and Distributed Training}} \\
Base model & Qwen3.8-27B \\
Training framework & SLIME / Megatron \\
Training strategy & Full-parameter SFT \\
Number of GPUs & 64 \\
Tensor parallelism & 4 \\
Pipeline parallelism & 2 \\
Context parallelism & 2 \\
Sequence parallelism & Enabled \\
\midrule
\multicolumn{2}{l}{\textit{Sequence Configuration}} \\
Global batch size & 128 \\
Maximum context length & 262,145 \\
Dynamic packing limit per GPU & 49,152 \\
\midrule
\multicolumn{2}{l}{\textit{Optimization}} \\
Optimizer & Adam \\
Adam $(\beta_1,\beta_2)$ & $(0.9,0.98)$ \\
Initial learning rate & $1\times10^{-5}$ \\
Minimum learning rate & $1\times10^{-6}$ \\
Learning rate schedule & Cosine decay \\
Warmup ratio & 0.05 \\
Weight decay & 0 \\
Loss function & Sample-level SFT loss \\
Loss masking & Assistant tokens only \\
\bottomrule
\end{tabular}
\end{table}

\subsection{Implementation and Benchmark Artifacts}

\paragraph{Implementation Audit.}
Reference outputs in SWR are generated by executable implementations of the corresponding software workflows. We conduct an abstract syntax tree (AST)-based audit of 580 source files and record file sizes and SHA-256 checksums for 815 reference files. Automated validation examines software dependency consistency and the validity of categorical, numerical, and structured reference outputs.

\paragraph{Benchmark Artifacts.}
The complete SWR100 package contains 2,819 files totaling 426\,MB, including 374\,MB of reference artifacts. Content-based deduplication yields 404 distinct blobs with a total size of 244\,MB. The package includes the domain-specific assets required for reference workflow execution. For instance, the URDF workflow contains the MuJoCo Menagerie UR5e XML model and 20 associated mesh assets, occupying approximately 33\,MB.

\FloatBarrier

\section{Dataset Composition and Quality Analysis}
\label{app:dataset-audit}

\subsection{Dataset Composition}

\begin{figure}[H]
\centering
\includegraphics[width=0.64\linewidth]{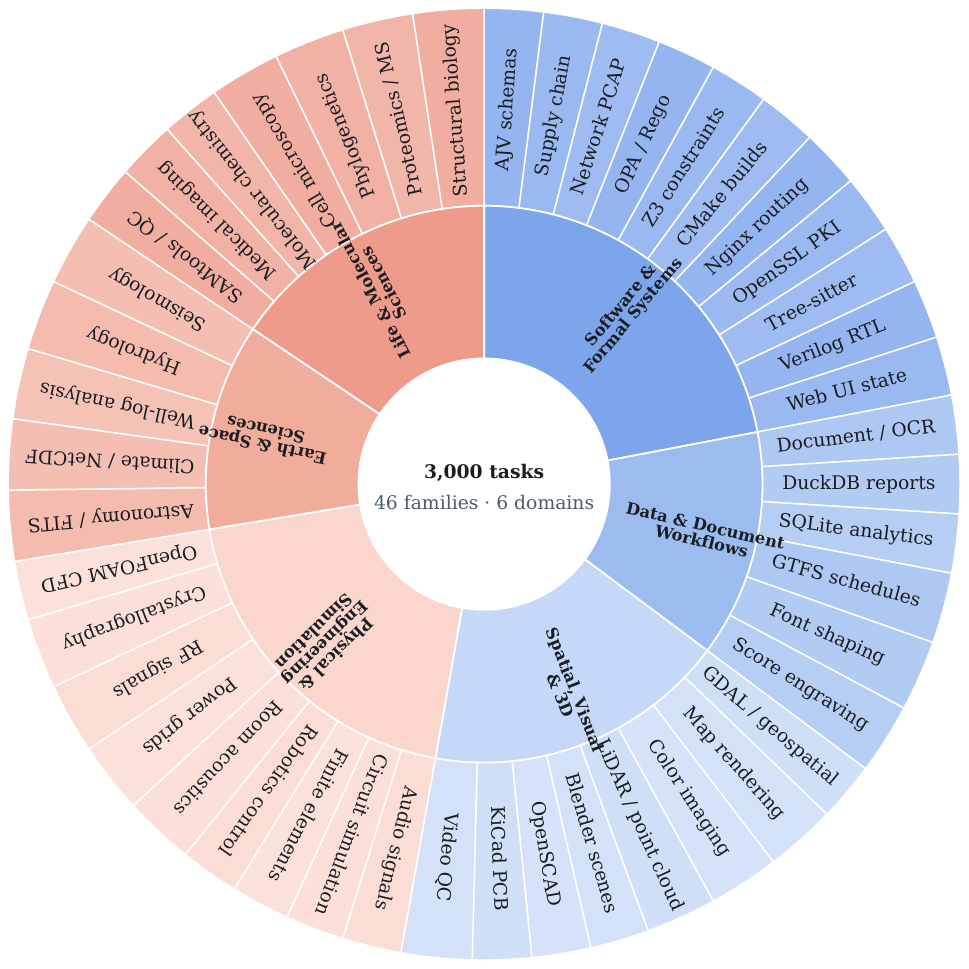}
\caption{Hierarchical composition of SWR across six application domains and 46 software families. No individual software family accounts for more than 2.4\% of the dataset.}
\label{fig:family-taxonomy}
\end{figure}

\paragraph{Domain and Software-Family Distribution.}
SWR comprises 3,000 reconstruction tasks derived from 500 reference workflows across 46 software families and six application domains: Software \& Formal Systems (11 families, 660 tasks), Physical \& Engineering Simulation (9 families, 588 tasks), Spatial, Visual \& 3D (8 families, 528 tasks), Life \& Molecular Sciences (7 families, 468 tasks), Data \& Document Workflows (6 families, 396 tasks), and Earth \& Space Sciences (5 families, 360 tasks). The domain-level distribution ranges from 12.0\% to 22.0\% of the dataset, with no individual software family accounting for more than 2.4\% of the tasks. Figure~\ref{fig:family-taxonomy} presents the hierarchical organization of the 46 software families.

\paragraph{Input Parameter Distribution.}
Each reconstruction task is associated with a typed input schema specifying the configurable parameters of its reference workflow. Approximately 86.7\% of tasks contain five input parameters, while tasks with three or four parameters each account for approximately 6.7\% of the dataset. Across all input parameters, numerical values constitute the largest proportion (66.1\%), followed by strings (26.7\%), Booleans (5.2\%), arrays (1.9\%), and mixed-type parameters (0.1\%). These parameters specify heterogeneous computational conditions, including numerical settings, configuration options, and categorical constraints.

\begin{figure*}[p]
\centering
\includegraphics[width=\textwidth]{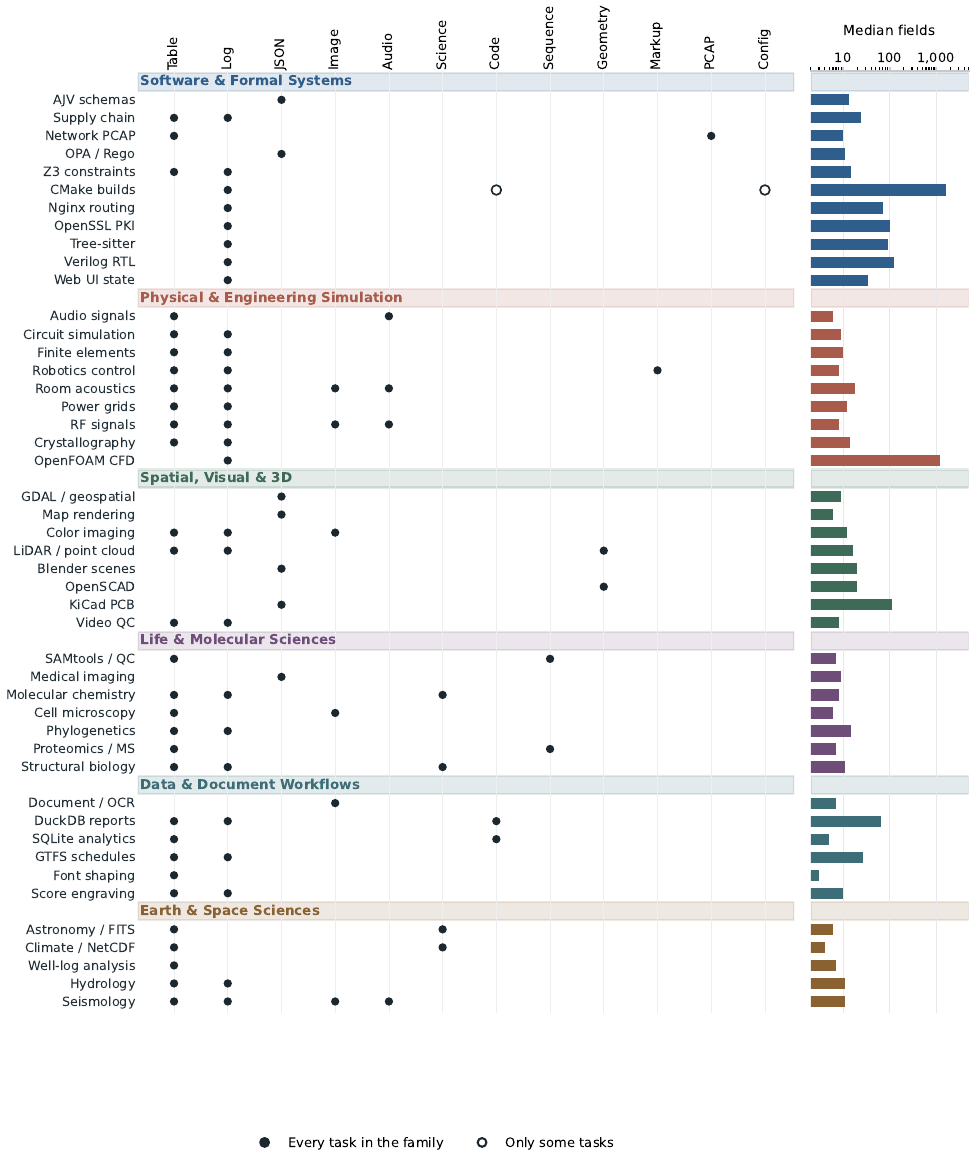}
\caption{Distribution of output modalities and verifiable output fields across the 46 software families. Filled markers indicate modalities present in all tasks within a family, whereas open markers indicate modalities present in a subset of tasks. For CMake, source code and configuration files appear in 36 and 54 of its 60 tasks, respectively. The horizontal bars indicate the median number of verifiable output fields per family on a logarithmic scale.}
\label{fig:family-modalities}
\end{figure*}

\paragraph{Output Modalities.}
SWR encompasses a diverse set of computational outputs, with individual tasks potentially involving multiple output modalities. Tabular outputs and textual outputs are the most prevalent, occurring in 2,028 and 1,776 tasks, respectively. Other output modalities include structured JSON (432), images (420), audio (276), scientific data (276), source code (156), biological sequences (132), geometry (132), markup (60), packet captures (60), and configuration files (54). Figure~\ref{fig:family-modalities} presents the distribution of output modalities across software families.

\begin{figure}[H]
\centering
\includegraphics[width=0.52\linewidth]{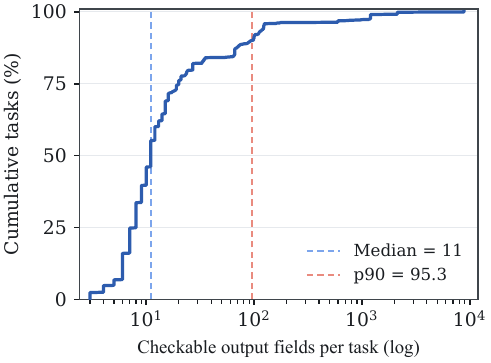}
\caption{Distribution of the number of verifiable output fields per reconstruction task in SWR, excluding digest-like profiling fields.}
\label{fig:semantic-complexity}
\end{figure}

\paragraph{Output Fields and Verification Requirements.}
We define an output field as an individual data item in the reference output that is subject to verification, such as a numerical quantity in a physical simulation, a categorical outcome in software execution, or a geometric property of a reconstructed object. A single output artifact may contain multiple such fields, each evaluated according to its corresponding comparison rule. As shown in Figure~\ref{fig:semantic-complexity}, after excluding digest-like profiling fields, the number of verifiable output fields per task has a median of 11 and a 90th percentile of 95.3. Tasks involving nested tables, geometric descriptors, and scientific arrays typically contain more output fields. Strict reconstruction success requires agreement with the reference outputs across all required fields and hidden scenarios, rather than correctness on a subset of the computational results.

\subsection{Input Sensitivity and Replay Analysis}

\paragraph{Input--Output Sensitivity.}
We examine whether variations in input configurations induce changes in the reference outputs at three levels: task-level sensitivity, field-level sensitivity, and parameter-level sensitivity.

\begin{itemize}
    \item \textbf{Task-Level Sensitivity.} All 3,000 tasks exhibit changes in their required outputs across at least one pair of input scenarios. For 2,984 tasks, every observed scenario pair produces different reference outputs. We compute scenario-pair sensitivity by averaging the per-task fraction of scenario pairs with different reference outputs, excluding digest-like fields.

    \item \textbf{Field-Level Sensitivity.} On average, 75.1\% of the required output fields exhibit variation across the sampled scenarios of a task, and 66.3\% of output fields differ between a pair of scenarios.

    \item \textbf{Parameter-Level Sensitivity.} For the 2,400 tasks supporting individual parameter perturbations, 89.5\% of input parameters induce measurable changes in the reference outputs when varied independently.
\end{itemize}

These statistics characterize input--output dependencies at complementary levels. Task-level sensitivity establishes that reference outputs are not invariant to input configurations, whereas field-level and parameter-level sensitivity quantify the extent to which these dependencies are reflected in individual output fields and configurable parameters.

\paragraph{Public-Output Replay Analysis.}
To quantify the extent to which hidden reference outputs can be approximated through direct reuse of public observations, we evaluate the three replay controls introduced in Appendix~\ref{app:specification}. Table~\ref{tab:shortcut-controls} reports both hidden-field agreement and strict reconstruction success. Most-frequent-public-output replay and nearest-input replay achieve 34.3\% and 38.0\% hidden-field agreement, respectively. Even with privileged access to hidden reference outputs for selecting public observations, oracle-selected best-public replay achieves only 44.3\% hidden-field agreement.

Although partial output agreement can be obtained through replay, none of the three controls satisfies the strict success criterion on any retained task. This zero success rate follows from the task acceptance criterion in ~\eqref{eq:task-validity}. The partial agreement scores further characterize the extent to which hidden outputs can be approximated using public observations, without implying resistance to all possible shortcut strategies.

\begin{table}[H]
\caption{Performance of public-output replay controls on SWR. Hidden-field agreement measures partial agreement with the hidden reference outputs, whereas strict success follows ~\eqref{eq:strict-success}.}
\label{tab:shortcut-controls}
\centering
\small
\setlength{\tabcolsep}{8pt}
\begin{tabular}{lrr}
\toprule
Method & Hidden-Field Agreement & Strict Success \\
\midrule
Reference workflow & 100.0\% & 100.0\% \\
\midrule
Most-frequent-output replay & 34.3\% & 0.0\% \\
Nearest-input replay & 38.0\% & 0.0\% \\
Oracle-selected replay & 44.3\% & 0.0\% \\
\bottomrule
\end{tabular}
\end{table}

\subsection{Within-Workflow Variant Analysis}

\paragraph{Scenario Overlap.}
Each reference workflow generates six reconstruction tasks with different input configurations. As a post-generation audit, we compare all 7,500 pairs of variants across the 500 workflows to examine potential scenario overlap. We find that no hidden input of a task appears in the public or hidden examples of another task derived from the same workflow. Moreover, the six tasks generated from each workflow have distinct sets of hidden reference outputs.

\paragraph{Behavioral Diversity.}
We further examine whether task variants differ only in input values or also in the computational behaviors to be reconstructed. Among 100 sampled workflows (600 tasks), all 100 contain output fields that remain constant in one task but vary with the inputs in another. In 79 workflows, at least one task produces a categorical outcome not observed in the other five tasks. In 90 workflows, at least one input parameter is fixed in one task but varied in another. For example, tasks derived from an OpenSSL workflow may require reconstructing certificate acceptance, rejection reasons, or extended-key-usage constraints, whereas tasks derived from a CMake workflow may exercise project configuration, installation, or package discovery. These results indicate that task variants derived from the same workflow can cover different computational behaviors, rather than merely different numerical inputs.

\FloatBarrier

\section{Reconstruction Performance Analysis}
\label{app:frozen-diagnostics}

\paragraph{Reconstruction Failure Analysis.}
We analyze why agents fail to reconstruct the reference workflows even when their submitted programs are executable. Among the candidate programs from 2,781 evaluated workspaces, 1,591 produce identical outputs for different hidden inputs, and 75 directly reproduce outputs from the public examples. These results reveal a major failure mode: agents generate executable programs without correctly modeling the relationship between workflow inputs and outputs. Although unsuccessful reconstructions achieve an average diagnostic agreement of 75.4\%, they fail to reproduce all required outputs across the hidden scenarios.

\begin{figure}[H]
\centering
\begin{subfigure}[t]{0.48\textwidth}
\centering
\includegraphics[width=\linewidth]{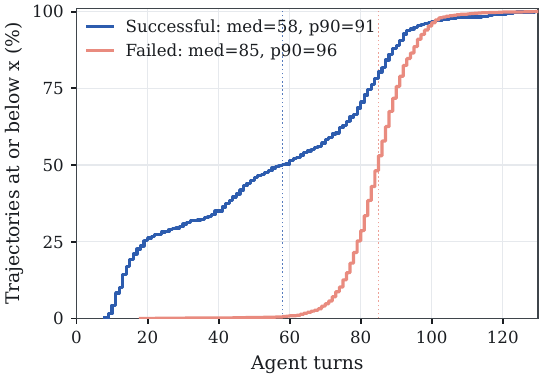}
\caption{Empirical cumulative distributions of interaction length for successful and model-attributable failed reconstructions.}
\label{fig:trajectory-length}
\end{subfigure}\hfill
\begin{subfigure}[t]{0.48\textwidth}
\centering
\includegraphics[width=\linewidth]{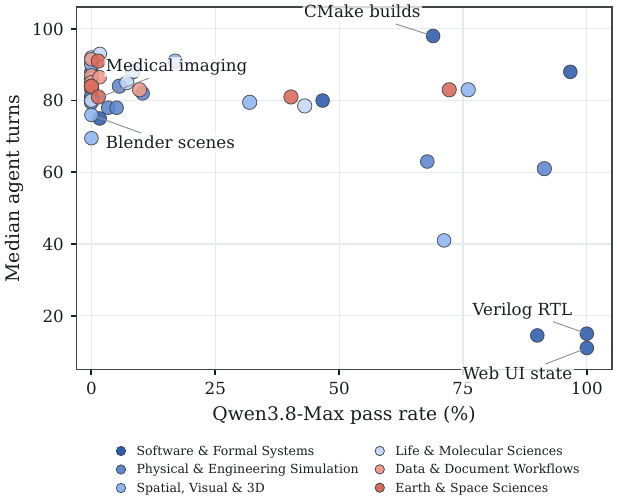}
\caption{Software-family success rate versus median interaction length.}
\label{fig:family-effort}
\end{subfigure}
\caption{Interaction length and reconstruction success in the single-attempt SWR evaluation using the finalized verifier.}
\label{fig:interaction-diagnostics}
\end{figure}

\paragraph{Interaction Length.}
We examine the relationship between reconstruction success and the number of interaction turns required by the agent. Successful reconstructions terminate after a median of 58 turns, whereas runs that fail due to model errors require a median of 85 turns. As shown in Figure~\ref{fig:trajectory-length}, this difference is consistently observed across the distributions of successful and unsuccessful trajectories, rather than being driven by a small number of exceptionally long runs. We further examine this relationship at the software-family level. Figure~\ref{fig:family-effort} reveals a negative correlation between reconstruction success rate and median interaction length ($r=-0.647$), indicating that agents generally spend more turns on software families with lower reconstruction success rates. These results suggest that longer interactions do not necessarily lead to successful reconstruction, particularly for workflows whose underlying computational behavior is difficult for the agent to recover.

\section{Reconstruction SFT Data Construction}
\label{app:sft-corpus}

\paragraph{Trajectory Selection and Sampling.}
The reconstruction corpus contains 1,422 unique verified trajectories. Trajectories associated with SWR100 task instances are excluded from SFT. We sample verified trajectories with repetition to construct the final reconstruction-only corpus of 3,000 training entries after preprocessing and length filtering.

\paragraph{Trajectory Preprocessing.}
Within each trajectory, we remove duplicate events and actions recorded after solution completion, and repair conversation boundaries. This cleanup preserves intentional repetition of complete trajectories across training entries.

\paragraph{Sequence Length Filtering.}
For the final SFT run initialized from the base model, we apply the chat template independently to each trajectory. The 49,152-token limit instead governs dynamic packing per GPU. The remaining optimization settings are specified in Appendix~\ref{app:specification}.

\section{Additional SFT Experiments}
\label{app:sft-pilot}

\subsection{Training-Trajectory Scaling and Corpus Comparisons}

\paragraph{Training-Trajectory Scaling.}
Table~\ref{tab:sft-seeds} reports the individual results underlying Table~\ref{tab:sft-results} for seeds 42, 43, and 44. The 3,000-entry condition improves over the base model on all four benchmarks for every seed, whereas improvements at smaller training-entry budgets vary across benchmarks and seeds.

\begin{table}[H]
\caption{Per-seed results for Qwen3.8-27B under different training-entry budgets. All scores are percentages; LHTB reports dense reward and the remaining benchmarks report task success.}
\label{tab:sft-seeds}
\centering
\small
\begin{tabular}{llrrrr}
\toprule
Model & Seed & TB2 (\%) & TB4 (\%) & LHTB & SWR100 \\
\midrule
Base & 42 & 47.19 & 0.00 & 22.00 & 1.00 \\
Base & 43 & 49.44 & 1.52 & 20.00 & 1.00 \\
Base & 44 & 47.19 & 0.00 & 20.00 & 2.00 \\
SFT-750 & 42 & 50.56 & 1.52 & 23.00 & 1.00 \\
SFT-750 & 43 & 49.44 & 1.52 & 22.00 & 2.00 \\
SFT-750 & 44 & 50.56 & 0.00 & 22.00 & 1.00 \\
SFT-1500 & 42 & 51.69 & 1.52 & 24.00 & 2.00 \\
SFT-1500 & 43 & 50.56 & 1.52 & 23.00 & 3.00 \\
SFT-1500 & 44 & 49.44 & 3.03 & 26.00 & 3.00 \\
SFT-3000 & 42 & 53.93 & 3.03 & 28.00 & 3.00 \\
SFT-3000 & 43 & 52.81 & 4.55 & 27.00 & 4.00 \\
SFT-3000 & 44 & 53.93 & 3.03 & 28.00 & 3.00 \\
\bottomrule
\end{tabular}
\end{table}

\paragraph{Matched-Token Corpus Comparison.}
For each of the four alternative terminal-agent corpora in Table~\ref{tab:sft-corpus-controls}, we fine-tune Qwen3.8-27B with 88.48M assistant-loss tokens, matching the number of tokens contributing to the SFT objective. Table~\ref{tab:sft-control-seeds} reports the individual evaluation results for the three seeds.

\begin{table}[H]
\caption{Per-seed results for Qwen3.8-27B trained on four alternative corpora with 88.48M assistant-loss tokens per run. All scores are percentages.}
\label{tab:sft-control-seeds}
\centering
\small
\begin{tabular}{llrrrr}
\toprule
Training corpus & Seed & TB2 & TB4 & LHTB & SWR100 \\
\midrule
OpenThoughts-Agent-v1 & 42 & 48.31 & 0.00 & 21.00 & 1.00 \\
OpenThoughts-Agent-v1 & 43 & 50.56 & 0.00 & 23.00 & 2.00 \\
OpenThoughts-Agent-v1 & 44 & 48.31 & 0.00 & 24.00 & 2.00 \\
TMax & 42 & 49.44 & 0.00 & 21.00 & 2.00 \\
TMax & 43 & 47.19 & 1.52 & 19.00 & 1.00 \\
TMax & 44 & 51.69 & 1.52 & 22.00 & 1.00 \\
LiteCoder-Terminal & 42 & 50.56 & 0.00 & 22.00 & 2.00 \\
LiteCoder-Terminal & 43 & 48.31 & 0.00 & 24.00 & 0.00 \\
LiteCoder-Terminal & 44 & 47.19 & 1.52 & 23.00 & 1.00 \\
Nemotron-Terminal & 42 & 48.31 & 0.00 & 23.00 & 1.00 \\
Nemotron-Terminal & 43 & 47.19 & 0.00 & 23.00 & 0.00 \\
Nemotron-Terminal & 44 & 46.07 & 1.52 & 25.00 & 3.00 \\
\bottomrule
\end{tabular}
\end{table}

\subsection{Controlled Comparisons and Transfer}
\label{app:controlled-studies}

\paragraph{Controlled Trajectory Collection.}
To control for differences in teacher capabilities and collection protocols, we collect three trajectory sets using Qwen3.8-Max-0902 and Terminus-2 with a budget of 100 turns and one hour per rollout: verified SWR trajectories, conventional terminal-task trajectories from TMax-15K, and failed SWR trajectories. TMax-15K contains newly collected rollouts and is distinct from the TMax corpus used in Table~\ref{tab:sft-corpus-controls}. All three conditions use full-parameter SFT with the same student configuration, 88.48M assistant-loss tokens, a global batch size of 128, a learning rate decaying from $10^{-5}$ to $10^{-6}$, a warmup ratio of 0.05, and seeds 42, 43, and 44.

As shown in Table~\ref{tab:same-teacher-control}, verified SWR trajectories outperform TMax-15K on TB2 and TB4 by 4.50 and 1.52 percentage points, respectively. Training on failed SWR trajectories yields a TB2 score of 46.82\%, below the base model (47.94\%). The advantage of verified SWR trajectories therefore persists under matched teacher, harness, rollout, and training-token budgets.

\begin{table}[H]
\caption{Controlled comparison of trajectory sources under matched teacher, harness, rollout limits, student configuration, and training-token budget (88.48M assistant-loss tokens). The base model receives no SFT.}
\label{tab:same-teacher-control}
\centering
\small
\begin{tabular}{lrrr}
\toprule
Training data & Seed & TB2 (\%) & TB4 (\%) \\
\midrule
Base & 42 & 47.19 & 0.00 \\
& 43 & 49.44 & 1.52 \\
& 44 & 47.19 & 0.00 \\
Base mean & -- & $47.94 \pm 1.30$ & $0.51 \pm 0.87$ \\
\midrule
SWR verified & 42 & 53.93 & 3.03 \\
& 43 & 52.81 & 4.55 \\
& 44 & 53.93 & 3.03 \\
SWR mean & -- & $\mathbf{53.56 \pm 0.65}$ & $\mathbf{3.54 \pm 0.87}$ \\
\midrule
TMax-15K & 42 & 47.19 & 3.03 \\
& 43 & 49.44 & 3.03 \\
& 44 & 50.56 & 0.00 \\
TMax-15K mean & -- & $49.06 \pm 1.72$ & $2.02 \pm 1.75$ \\
\midrule
Failed SWR & 42 & 47.19 & 0.00 \\
& 43 & 46.07 & 1.52 \\
& 44 & 47.19 & 1.52 \\
Failed SWR mean & -- & $46.82 \pm 0.65$ & $1.01 \pm 0.87$ \\
\bottomrule
\end{tabular}
\end{table}

\paragraph{Trajectory Diversity and Training Exposure.}
We first vary the number of unique verified trajectories while fixing the assistant-loss-token budget at 88.48M. Increasing the number of unique trajectories from 250 to 838 improves TB2 from 48.31\% to 50.56\% (Table~\ref{tab:unique-trajectory-control}), indicating a benefit from greater trajectory diversity under a fixed token budget.

We further compare training once on all 1,422 unique trajectories with repeated sampling from the same set. The two conditions consume 41.92M and 88.48M assistant-loss tokens and achieve TB2 scores of 48.69\% and 53.56\%, respectively.

\begin{table}[H]
\caption{Effects of unique trajectory count and repeated sampling on TB2. The unique-only condition uses a smaller training-token budget. Results are mean and standard deviation over three seeds.}
\label{tab:unique-trajectory-control}
\centering
\small
\begin{tabular}{lrrr}
\toprule
Setting & Unique trajectories & Assistant-loss tokens & TB2 Pass@1 \\
\midrule
Base & -- & -- & $47.94 \pm 1.30$ \\
Subset-250 & 250 & 88.48M & $48.31 \pm 1.12$ \\
Subset-500 & 500 & 88.48M & $49.06 \pm 1.72$ \\
Subset-838 & 838 & 88.48M & $50.56 \pm 1.12$ \\
Unique-only, no repetition & 1,422 & 41.92M & $48.69 \pm 0.65$ \\
Full SWR, repeated sampling & 1,422 & 88.48M & $\mathbf{53.56 \pm 0.65}$ \\
\bottomrule
\end{tabular}
\end{table}

\paragraph{Transfer to Software Families Absent from SFT.}
To examine whether reconstruction SFT improves the ability of agents to reconstruct workflows from software families not represented in the training corpus, we identify 20 software families for which the teacher fails to produce any successful reconstruction trajectories. Since none of these families contributes training data to reconstruction SFT, the corresponding 46 tasks in SWR100 provide an opportunity to evaluate transfer across software families. As shown in Table~\ref{tab:unseen-family-transfer}, reconstruction SFT improves the mean success rate from $1.45\pm1.26$\% to $4.35\pm2.17$\% across three independent runs. At the domain level, the success rate increases from 4.76\% to 14.29\% in Software \& Formal Systems and from 16.67\% to 33.33\% in Physical \& Engineering Simulation. Reconstruction SFT also enables successful reconstruction in Spatial, Visual \& 3D, where the base model fails on all tasks across the three runs. Neither model successfully reconstructs tasks in the remaining three domains. These results suggest that the benefits of reconstruction SFT are not restricted to the software families represented in the training corpus.

\begin{table}[H]
\caption{Reconstruction performance on 46 SWR100 tasks from 20 software families absent from reconstruction SFT. Results are mean and standard deviation (\%) over three seeds.}
\label{tab:unseen-family-transfer}
\centering
\scriptsize
\setlength{\tabcolsep}{4pt}
\begin{tabular}{lrrr}
\toprule
Domain & Tasks & Base & Reconstruction SFT \\
\midrule
Software \& Formal Systems & 7 & $4.76 \pm 8.25$ & $14.29 \pm 0.00$ \\
Physical \& Engineering Simulation & 2 & $16.67 \pm 28.87$ & $33.33 \pm 28.87$ \\
Spatial, Visual \& 3D & 16 & $0.00 \pm 0.00$ & $2.08 \pm 3.61$ \\
Life \& Molecular Sciences & 9 & $0.00 \pm 0.00$ & $0.00 \pm 0.00$ \\
Data \& Document Workflows & 6 & $0.00 \pm 0.00$ & $0.00 \pm 0.00$ \\
Earth \& Space Sciences & 6 & $0.00 \pm 0.00$ & $0.00 \pm 0.00$ \\
\midrule
Overall & 46 & $1.45 \pm 1.26$ & $\mathbf{4.35 \pm 2.17}$ \\
\bottomrule
\end{tabular}
\end{table}

The unseen-family task counts are CMake (3), OpenSSL (3), Network PCAP (1), OpenFOAM (2), Blender (4), KiCad (3), GDAL (3), map rendering (3), LiDAR (2), OpenSCAD (1), molecular chemistry (3), cell microscopy (3), structural biology (2), medical imaging (1), font shaping (3), DuckDB (2), SQLite (1), well-log (3), climate (2), and seismology (1).

\paragraph{Transfer Across Student Models.}
We repeat reconstruction SFT with Qwen3.5-27B to assess whether the observed gains extend to a different student version. As reported in Table~\ref{tab:second-student}, SFT improves TB2 by 6.00 percentage points for Qwen3.5-27B, compared with 5.62 percentage points for Qwen3.8-27B.

\begin{table}[H]
\caption{TB2 Pass@1 (\%) before and after reconstruction SFT for two student models. Results are mean and standard deviation over three seeds.}
\label{tab:second-student}
\centering
\small
\begin{tabular}{lrr}
\toprule
Student & Base & Reconstruction SFT \\
\midrule
Qwen3.8-27B & $47.94\pm1.30$ & $53.56\pm0.65$ \\
Qwen3.5-27B & $41.57\pm1.12$ & $47.57\pm1.72$ \\
\bottomrule
\end{tabular}
\end{table}

\paragraph{Number of Variants per Workflow.}
Using the same 50 reference workflows across six domains, we compare six and ten task variants per workflow, yielding 300 and 500 tasks, respectively. Four independent teacher rollouts per task produce 1,200 and 2,000 rollouts. Each SFT condition uses 1,000 successful trajectory entries, with repetition where necessary. Increasing the number of task variants from six to ten, together with the corresponding expansion of trajectory collection, improves TB2 by 0.75 percentage points under a fixed budget of 1,000 training entries (Table~\ref{tab:variant-scaling}).

\begin{table}[H]
\caption{Effect of task-variant count for a fixed set of 50 workflows. Each task receives four teacher rollouts. SFT entries may include repeated trajectories. Collection budgets differ across conditions.}
\label{tab:variant-scaling}
\centering
\small
\begin{tabular}{rrrrr}
\toprule
Variants/workflow & Tasks & Teacher rollouts & SFT entries & TB2 Pass@1 (\%) \\
\midrule
6 & 300 & 1,200 & 1,000 & $48.31\pm2.25$ \\
10 & 500 & 2,000 & 1,000 & $49.06\pm0.65$ \\
\bottomrule
\end{tabular}
\end{table}

\section{Model Performance on SWR100}
\label{app:swr100-results}

\paragraph{Evaluation Coverage.}
SWR100 contains 19 tasks in Software \& Formal Systems, 20 in Physical \& Engineering Simulation, 19 in Spatial, Visual \& 3D, 15 in Life \& Molecular Sciences, 14 in Data \& Document Workflows, and 13 in Earth \& Space Sciences. Each of the eight models is evaluated once on the same 100 tasks. Of the 800 scheduled trials, 797 have execution records, 785 receive verifier scores, and 44 satisfy the strict success criterion. Missing and unscored trials remain in the denominator when computing Pass@1.

\paragraph{Model and Domain Performance.}
GPT-5.6 Sol solves 20 tasks, compared with six for Kimi-K3, four for Grok~4.6, and two to three for each remaining model. Across all models, 27 of the 44 successful trials occur in Software \& Formal Systems, eight in Physical \& Engineering Simulation, eight in Spatial, Visual \& 3D, and one in Life \& Molecular Sciences. No model solves a task in Data \& Document Workflows or Earth \& Space Sciences under the reported evaluation protocol.

\section{Interaction Efficiency and Statistical Analysis}
\label{app:swr100-diagnostics}

\begin{figure}[H]
\centering
\begin{subfigure}[t]{0.49\textwidth}
\centering
\includegraphics[width=\linewidth]{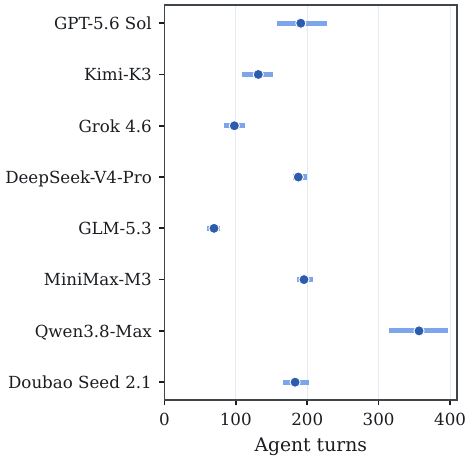}
\caption{Interaction length (agent turns).}
\label{fig:swr100-interaction-length}
\end{subfigure}\hfill
\begin{subfigure}[t]{0.49\textwidth}
\centering
\includegraphics[width=\linewidth]{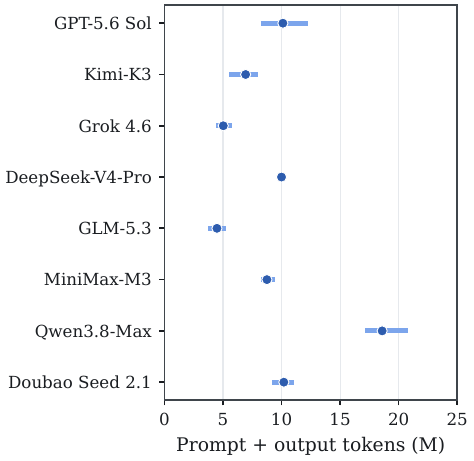}
\caption{Token consumption (prompt and output tokens).}
\label{fig:swr100-tokens}
\end{subfigure}
\caption{Distribution of agent interaction lengths and token consumption across foundation models on SWR100 under a three-hour time budget per task. Points represent the median across evaluated tasks, and horizontal bars indicate the 25th to 75th percentile range.}
\label{fig:swr100-trajectories}
\end{figure}

\paragraph{Interaction Length and Token Consumption.}
We analyze the interaction patterns of different foundation models on SWR100 in terms of agent turns and token consumption. Despite identical interaction limits, models exhibit substantial differences in their utilization of the available budget. As shown in Figure~\ref{fig:swr100-trajectories}, median interaction length ranges from 69.5 to 356.5 turns, while median token consumption ranges from 4.49M to 18.59M. Qwen3.8-Max records the highest median interaction length and token consumption (356.5 turns and 18.59M tokens), whereas GLM-5.3 records the lowest (69.5 turns and 4.49M tokens). Despite these substantial differences in resource consumption, both models achieve a reconstruction success rate of 3.0\%. GPT-5.6 Sol achieves the highest success rate of 20.0\% with a median of 191 turns and 10.12M tokens, substantially below the interaction expenditure of Qwen3.8-Max. These results reveal a discrepancy between interaction expenditure and reconstruction performance, suggesting that the ability to effectively utilize computational resources is critical to scientific workflow reconstruction. Detailed per-model statistics are presented in Table~\ref{tab:swr100-details}.

\paragraph{Comparative Model Performance.}
Under the single-attempt evaluation protocol, GPT-5.6 Sol achieves a success rate of 20.0\%, outperforming the second-best model, Kimi-K3, by 14 percentage points. All seven other models achieve success rates between 2.0\% and 6.0\%, despite substantial differences in their interaction lengths and token consumption. The performance distribution reveals a pronounced gap between the strongest model and the other evaluated models, while the latter exhibit uniformly low reconstruction success rates. Given that SWR100 requires candidate programs to reproduce workflow outputs across multiple hidden input configurations, these results highlight the difficulty of recovering input--output dependencies in specialized computational workflows. SWR100 therefore provides a challenging evaluation setting for examining the capabilities of foundation models in scientific computation and executable program reconstruction. Each reported score is obtained from one attempt per model--task pair under the same evaluation budget.

\begin{table}[H]
\caption{Per-model evaluation coverage, strict successes, and interaction expenditure on SWR100. Each model is scheduled on 100 tasks. Turns and token consumption are reported as medians.}
\label{tab:swr100-details}
\centering
\small
\setlength{\tabcolsep}{3.5pt}
\begin{tabular}{lrrrrr}
\toprule
Model & Records & Scored & Pass & Turns & Tokens (M) \\
\midrule
GPT-5.6 Sol & 99 & 89 & 20 & 191.0 & 10.12 \\
Kimi-K3 & 98 & 98 & 6 & 131.5 & 6.94 \\
Grok 4.6 & 100 & 99 & 4 & 98.0 & 5.03 \\
DeepSeek-V4-Pro & 100 & 99 & 3 & 187.5 & 10.00 \\
Qwen3.8-Max & 100 & 100 & 3 & 356.5 & 18.59 \\
GLM-5.3 & 100 & 100 & 3 & 69.5 & 4.49 \\
MiniMax-M3 & 100 & 100 & 3 & 195.5 & 8.75 \\
Doubao Seed 2.1 & 100 & 100 & 2 & 183.0 & 10.20 \\
\midrule
Total & 797 & 785 & 44 & -- & -- \\
\bottomrule
\end{tabular}
\end{table}

\section{Verifier Reliability and Ablation Studies}
\label{app:verifier-stress}

\subsection{Verifier Calibration}

\paragraph{Mutation Protocol.}
We evaluate the released task comparators using controlled modifications of reference outputs. The structured-output tests comprise five mutation families:
\begin{itemize}
    \item \textbf{Numerical perturbation.} Perturb one finite floating-point leaf value by $0.5$, $0.9$, $0.99$, $1.01$, $1.1$, $2$, or $10$ times the configured numerical tolerance.
    \item \textbf{Deep-field modification.} Modify an output field at the deepest nesting level while preserving the surrounding schema.
    \item \textbf{Correlated numerical shift.} Apply a correlated shift to all numeric leaf values.
    \item \textbf{Scenario substitution.} Replace the target output with a reference output generated for another hidden scenario.
    \item \textbf{Ignored input parameter.} Replace one hidden input parameter with its value in a public scenario and rerun the reference workflow, producing a schema-valid output for an incorrect input configuration.
\end{itemize}
For geometric outputs, we test uniform scaling, localized surface deformation, facet deletion, and mesh substitution across public or hidden scenarios. Each trial invokes the task comparator directly. The 60 bioinformatics tasks are evaluated in their packaged \texttt{samtools}/\texttt{bcftools}/\texttt{seqkit} environment.

\begin{table}[H]
\caption{Verifier calibration results on the full SWR dataset. Correct verification requires accepting valid controls and rejecting all invalid output mutations.}
\label{tab:verifier-stress}
\centering
\small
\setlength{\tabcolsep}{5pt}
\begin{tabular}{lrrr}
\toprule
Mutation class & Trials & Accepted & Rejected \\
\midrule
Valid near-boundary controls & 7,065 & 7,065 & 0 \\
\midrule
Single deep-leaf change & 2,940 & 0 & 2,940 \\
Correlated numeric shift & 2,868 & 0 & 2,868 \\
Wrong hidden scenario & 2,939 & 0 & 2,939 \\
Ignored input parameter & 9,901 & 0 & 9,901 \\
Numeric perturbation $>1\times$ tolerance & 9,260 & 0 & 9,260 \\
Localized geometry deformation & 240 & 0 & 240 \\
Facet deletion & 300 & 0 & 300 \\
Wrong-scenario geometry & 120 & 0 & 120 \\
Uniform scale $\geq1.02\times$ & 240 & 0 & 240 \\
\midrule
All invalid or wrong-scenario mutations & 28,808 & 0 & 28,808 \\
\bottomrule
\end{tabular}
\end{table}

\begin{figure*}[t]
\centering
\begin{subfigure}[t]{0.48\textwidth}
\centering
\includegraphics[width=\linewidth]{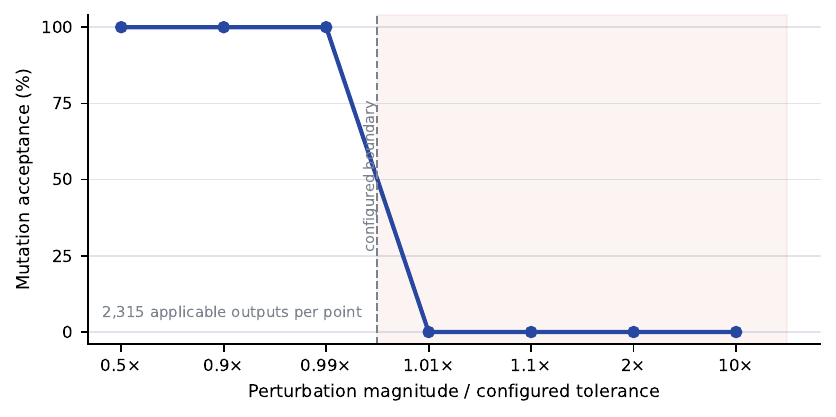}
\caption{Acceptance rate as a function of numerical perturbation relative to the configured tolerance.}
\label{fig:verifier-tolerance-boundary}
\end{subfigure}\hfill
\begin{subfigure}[t]{0.48\textwidth}
\centering
\includegraphics[width=\linewidth]{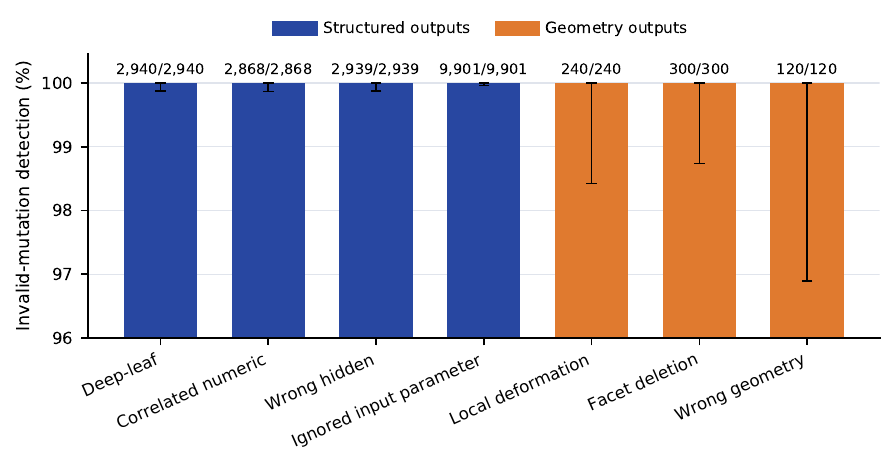}
\caption{Rejection rates for structured-output and geometry mutations.}
\label{fig:verifier-coherent-mutations}
\end{subfigure}
\caption{Verifier calibration results. (a) Numerical acceptance rates for 2,315 outputs with finite floating-point values at each perturbation magnitude. (b) Mutation rejection rates with Wilson 95\% lower confidence bounds; labels denote rejected and total cases.}
\label{fig:verifier-stress}
\end{figure*}

\paragraph{Calibration Results.}
Figure~\ref{fig:verifier-stress} and Table~\ref{tab:verifier-stress} summarize the verifier calibration results on 35,873 test cases. For numerical outputs, the verifier accepts all tested perturbations up to $0.99\times$ the configured tolerance and rejects those at $1.01\times$ and above. It also detects all tested modifications to nested output fields, correlated numerical shifts, wrong-scenario outputs, and 9,901 cases involving ignored input parameters. For geometric outputs, the verifier accepts equivalent representations produced through ASCII reserialization, facet reordering, and rigid translation, while rejecting local deformations, facet deletions, incorrect meshes, and scale changes of at least 2\%. Based on these experiments, we adopt family-specific relative tolerances for audio and geospatial outputs and incorporate bidirectional surface coverage and boundary-edge checks into geometric verification. Following calibration, the verifier accepts all 7,065 valid controls and rejects all 28,808 invalid mutations.

\subsection{Independent Verifier Audit}

\paragraph{Audit Protocol.}
To evaluate the reliability of the calibrated verifier, we construct an independent set of mutation cases that are not used during verifier calibration. We first validate the mutation generator on a pilot set of 100 tasks and subsequently fix both the mutation generation procedure and verifier thresholds. The final audit set comprises 100 distinct tasks excluded from the pilot set, including two tasks from each of the 46 software families and eight additional tasks selected for their challenging verification requirements. For these tasks, we construct 1,200 test cases, consisting of 400 valid output variations that preserve computational equivalence and 800 invalid output mutations. The invalid mutations introduce inconsistencies between related output fields, incorrect units, numerical boundary violations, combinations of outputs from different scenarios, and geometric distortions.

\paragraph{Audit Results.}
As shown in Table~\ref{tab:heldout-verifier-audit}, the calibrated verifier correctly classifies 1,191 of the 1,200 audit cases, achieving an overall agreement of 99.25\% with the predefined validity labels. For structured outputs, the verifier accepts all 376 valid cases and rejects 750 of the 752 invalid cases. For geometric outputs, it accepts 18 of the 24 valid cases and rejects 47 of the 48 invalid cases. Across all audit cases, the verifier incorrectly accepts three invalid outputs and rejects six valid outputs. The majority of disagreements arise in geometric verification, particularly when different geometric representations describe equivalent underlying surfaces.

\begin{table}[H]
\caption{Independent evaluation of the calibrated verifier on 1,200 audit cases. False acceptance and false rejection are determined relative to the predefined validity labels.}
\label{tab:heldout-verifier-audit}
\centering
\small
\setlength{\tabcolsep}{7pt}
\begin{tabular}{lrrrr}
\toprule
Audit Category & Cases & False Accept & False Reject & Agreement \\
\midrule
Structured valid outputs & 376 & 0 & 0 & 100.00\% \\
Structured invalid outputs & 752 & 2 & 0 & 99.73\% \\
Geometric valid outputs & 24 & 0 & 6 & 75.00\% \\
Geometric invalid outputs & 48 & 1 & 0 & 97.92\% \\
\midrule
Overall & 1,200 & 3 & 6 & 99.25\% \\
\bottomrule
\end{tabular}
\end{table}

\paragraph{Analysis of Verification Errors.}
Among the 1,200 audit cases, the verifier disagrees with the predefined validity labels in nine cases, including three false acceptances and six false rejections. The corresponding false-acceptance and false-rejection rates are 0.38\% (3/800) and 1.50\% (6/400), respectively. We examine these cases to identify the sources of disagreement. Two false acceptances involve numerical perturbations in circuit outputs, where the modified values differ from the reference values by one unit but remain within the configured relative tolerance. The remaining case involves a geometric shear that is not detected by the geometry comparator. All six false rejections occur when geometrically equivalent surfaces are represented using different mesh tessellations. Although these representations preserve the underlying surface geometry, they fail the sampled-point coverage check. These cases identify two principal challenges in verifying heterogeneous computational outputs: distinguishing meaningful numerical errors from permissible deviations and recognizing geometrically equivalent outputs with different discretizations.

\paragraph{Human Evaluation of Verifier Reliability.}
To independently assess the correctness of verifier decisions, we conduct a blinded human evaluation on 200 audit cases. For each of the 100 audited tasks, we select one output labeled valid and one labeled invalid. Two annotators independently examine whether each output is computationally equivalent to the corresponding reference output, without access to the predefined validity labels or verifier decisions. Disagreements between the annotators are resolved through adjudication.

As shown in Table~\ref{tab:heldout-human-audit}, the two annotators agree on 192 of the 200 cases (96.0\%). The predefined validity labels and verifier decisions each agree with the adjudicated human judgments on 196 cases (98.0\%). For the 184 cases sampled across all 46 software families, the verifier achieves 98.37\% agreement with human judgments. For the remaining 16 cases selected for their challenging verification requirements, the agreement reaches 93.75\%. These results demonstrate high consistency between automated verification and human judgments across heterogeneous computational outputs.

\begin{table}[H]
\caption{Human evaluation of verifier reliability on 200 audit cases. Predefined-label agreement and verifier agreement measure consistency with the adjudicated human judgments. Annotator agreement measures consistency between the two independent annotators before adjudication.}
\label{tab:heldout-human-audit}
\centering
\small
\setlength{\tabcolsep}{4pt}
\resizebox{\linewidth}{!}{
\begin{tabular}{lrrrrr}
\toprule
Audit Category & Tasks & Cases & Predefined-Label Agreement & Verifier Agreement & Annotator Agreement \\
\midrule
Software-family coverage & 92 & 184 & 98.91\% & 98.37\% & 96.74\% \\
High-risk verification cases & 8 & 16 & 87.50\% & 93.75\% & 87.50\% \\
\midrule
Overall & 100 & 200 & 98.00\% & 98.00\% & 96.00\% \\
\bottomrule
\end{tabular}
}
\end{table}

\paragraph{False Acceptance and False Rejection.}
We further analyze verifier decisions using the adjudicated human judgments as reference labels. Table~\ref{tab:heldout-adjudicated-confusion} presents the resulting confusion matrix. Among the 100 outputs judged valid by human annotators, the verifier correctly accepts 98 and incorrectly rejects two. Similarly, among the 100 outputs judged invalid, the verifier correctly rejects 98 and incorrectly accepts two. Both the false-acceptance and false-rejection rates are therefore 2.0\%. The balanced error rates indicate that the verifier maintains a high level of agreement with human judgments in both accepting computationally equivalent outputs and rejecting incorrect reconstructions. The remaining disagreements primarily concern numerical tolerance boundaries and geometric equivalence.

\begin{table}[H]
\caption{Comparison of automated verifier decisions with adjudicated human judgments on 200 audit cases.}
\label{tab:heldout-adjudicated-confusion}
\centering
\small
\setlength{\tabcolsep}{8pt}
\begin{tabular}{lrrrr}
\toprule
Human Judgment & Accepted & Rejected & Total & Agreement \\
\midrule
Valid outputs & 98 & 2 & 100 & 98.0\% \\
Invalid outputs & 2 & 98 & 100 & 98.0\% \\
\midrule
Overall & 100 & 100 & 200 & 98.0\% \\
\bottomrule
\end{tabular}
\end{table}

Taken together, the post-freeze audit and human evaluation establish complementary measures of verifier reliability. The former achieves 99.25\% agreement with predefined validity labels across 1,200 controlled test cases, while the latter achieves 98.0\% agreement with adjudicated human judgments across 200 cases. Both evaluations identify numerical tolerances and geometric equivalence as the primary sources of verification errors.

\subsection{Ablation of Task-Validation Components}

\paragraph{Experimental Setup.}
We investigate the contributions of two task-validation components: output comparison and public-output replay checks. Specifically, we remove each component independently during task generation and construct two alternative task sets. The former examines the importance of validating reference outputs, while the latter examines the importance of excluding tasks that can be solved by directly reusing public examples. For each resulting task set, we collect reconstruction trajectories and apply the same complete verifier to select successful trajectories for SFT. This ensures that the ablation modifies task validation while preserving the trajectory selection criterion across all conditions. We evaluate each condition in terms of task quality and downstream training performance. Task quality is measured using two metrics: (i) the invalid-task rate, defined as the proportion of tasks containing incorrect reference outputs or admitting shortcut solutions, and (ii) the shortcut success rate, defined as the proportion of tasks solved by the predefined public-output replay strategies. To evaluate the training utility of the resulting tasks, we fine-tune Qwen3.8-27B on the collected verified trajectories and report Pass@1 on TB2.

\paragraph{Effect of Task Validation.}
Table~\ref{tab:validation-ablation} presents the ablation results. Removing output comparison increases the number of accepted workflows from 500 to 558, yielding 3,348 tasks, of which 8.0\% are identified as invalid. The corresponding TB2 performance decreases from 53.56\% to 48.31\%. Removing public-output replay checks yields 3,252 tasks, with an invalid-task rate of 9.9\% and a shortcut success rate of 6.8\%. Training on trajectories collected from these tasks achieves 50.19\% on TB2. In contrast, the full pipeline produces 3,000 tasks with no invalid tasks or successful replay shortcuts identified by the quality audit and achieves the highest downstream performance.

\textbf{Both ablations increase the number of accepted tasks but reduce downstream training performance.} In particular, removing output comparison and public-output replay checks results in performance reductions of 5.25 and 3.37 percentage points on TB2, respectively. These results highlight the importance of task validity in scaling reconstruction-based training environments.

\begin{table}[H]
\caption{Ablation of task-validation components on dataset quality and downstream training performance. All conditions retain the complete verifier for trajectory selection. TB2 Pass@1 is reported as mean and standard deviation over three seeds.}
\label{tab:validation-ablation}
\centering
\small
\setlength{\tabcolsep}{4pt}
\resizebox{\linewidth}{!}{
\begin{tabular}{lrrrrr}
\toprule
Validation Setting & Workflows & Tasks & Invalid Tasks & Shortcut Success & TB2 Pass@1 (\%) \\
\midrule
Base (no SFT) & -- & -- & -- & -- & $47.94 \pm 1.30$ \\
\midrule
w/o Output Comparison & 558 & 3,348 & 8.0\% & 0.0\% & $48.31 \pm 1.12$ \\
w/o Public-Output Replay Checks & 542 & 3,252 & 9.9\% & 6.8\% & $50.19 \pm 1.72$ \\
\midrule
SWR & 500 & 3,000 & \textbf{0.0\%} & \textbf{0.0\%} & $\mathbf{53.56 \pm 0.65}$ \\
\bottomrule
\end{tabular}
}
\end{table}

\section{Reconstruction Failure Analysis}
\label{app:trajectory-case-studies}

\paragraph{Failure Classification.}
We analyze all 800 reconstruction attempts on SWR100 using the final verifier reports. Each attempt is assigned a primary outcome according to the order of verification checks: candidate program availability, execution on hidden inputs, shortcut detection, and output correctness. Output errors are further categorized into schema violations, numerical mismatches, categorical mismatches, and geometric discrepancies. We additionally record co-occurring output errors to characterize failures involving multiple types of discrepancies.

\paragraph{Overall Failure Patterns.}
Among the 800 reconstruction attempts, 44 satisfy the strict success criterion and 756 fail. Numerical mismatch is the dominant failure category, accounting for 456 unsuccessful attempts (60.3\%), followed by failure to produce an executable candidate program (130 attempts, 17.2\%). Other failure categories include categorical mismatches (55), schema violations (38), hidden execution or output-parsing errors (36), public-output replay (20), geometric mismatches (8), and missing records or residual outcomes (13). The prevalence of numerical mismatches indicates that generating executable programs alone is insufficient for scientific workflow reconstruction, where agents must accurately reproduce computational results across different input configurations.

\begin{figure}[H]
\centering
\includegraphics[width=\linewidth]{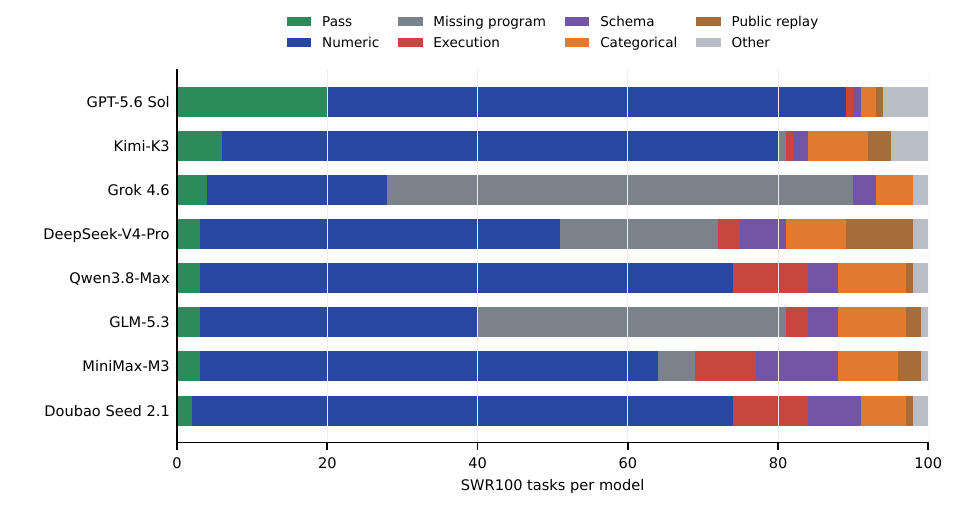}
\caption{Distribution of reconstruction outcomes across eight foundation models on SWR100. Each model is evaluated on the same 100 tasks. Primary failure categories are assigned according to the order of verification checks. Geometric mismatches, missing records, and other unscored outcomes are grouped into Other.}
\label{fig:swr100-failure-signatures}
\end{figure}

\paragraph{Model-Specific Failure Patterns.}
As shown in Figure~\ref{fig:swr100-failure-signatures}, the primary reconstruction bottlenecks vary substantially across models. Grok~4.6, GLM-5.3, and DeepSeek-V4-Pro account for 124 of the 130 missing-program failures, with 62, 41, and 21 cases, respectively. In contrast, numerical mismatch dominates the failures of GPT-5.6 Sol (69), Kimi-K3 (74), Qwen3.8-Max (71), and Doubao Seed~2.1 (72). These results reveal distinct limitations in program construction and computational accuracy across foundation models.

\paragraph{Co-occurring Output Errors.}
Numerical and categorical mismatches co-occur in 173 reconstruction attempts, categorical and structural errors in 22 attempts, and numerical and structural errors in 19 attempts. The prevalence of joint numerical and categorical errors indicates that workflow reconstruction requires recovering both numerical computations and the conditions governing discrete outputs.

\end{document}